\documentclass[reprint,tightenlines,superscriptaddress,prd,nofootinbib,eqsecnum,showpacs]{revtex4}
\usepackage{subfigure}
\usepackage{amssymb}
\usepackage{amsmath}
\usepackage[pdftex]{graphicx}
\def\_#1{^{}_{#1}}

\def\beq{\begin{equation}}
\def\eeq{\end{equation}}
\def\bea{\begin{eqnarray}}
\def\eea{\end{eqnarray}}
\usepackage{enumerate}

\begin{document}
\title{Anisotropic evolution of $D$-dimensional FRW spacetime}

\author{Chad Middleton}
\email{chmiddle@coloradomesa.edu}
\author{Bret A. Brouse Jr.}
\email{babrouse@mavs.coloradomesa.edu}
\author{Scott D. Jackson}
\email{sdjackson@mavs.coloradomesa.edu}

\affiliation{Department of Physical and Environmental Sciences, Colorado Mesa University, Grand Junction, Colorado 81501, USA}
\date{\today}

%
\begin{abstract}
We examine the time evolution of the $D=d+4$ dimensional Einstein field equations subjected to a flat Robertson-Walker metric where the 3D and higher-dimensional scale factors are allowed to evolve at different rates.  We find the exact solution to these equations for a single fluid component, which yields two limiting regimes offering the 3D scale factor as a function of the time.  The fluid regime solution closely mimics that described by 4D FRW cosmology, offering a late-time behavior for the 3D scale factor after becoming valid in the early universe, and can give rise to a late-time accelerated expansion driven by vacuum energy.  This is shown to be preceded by an earlier volume regime solution, which offers a very early-time epoch of accelerated expansion for a radiation-dominated universe for $d=1$.  The time scales describing these phenomena, including the transition from volume to fluid regime, are shown to fall within a small fraction of the first second when the fundamental constants of the theory are aligned with the Planck time.  This model potentially offers a higher-dimensional alternative to scalar-field inflationary theory and a consistent cosmological theory, yielding a unified description of early- and late-time accelerated expansions via a 5D spacetime scenario.
\end{abstract}

\maketitle

\section{Introduction}\label{intro}

In an attempt at unifying gravity and electromagnetism, Theodor Kaluza and Oskar Klein extended general relativity to a 5D spacetime ultimately hypothesizing the existence of an extra spatial dimension.\cite{KK} It was later shown that when the components of the 5D metric are chosen to be independent of the extra spatial dimension, the 5D vacuum field equations yield the 4D Einstein field equations of gravitation and Maxwell's equations for electromagnetic theory, plus an additional equation that determines the dynamics of a scalar field.  Since the advent of Kaluza-Klein theory, the notion of incorporating extra dimensions into 4D theories to potentially explain certain physical phenomena continues to garner attention and often falls under the designation of Modern Kaluza-Klein theories (for a review, see Applequist, Chodos, and Frend and/or Wesson \cite{Apple}).  Since the discovery of string theory in the late 1960s, research into extra dimensional scenarios has only increased in popularity, many years after Kaluza and Klein's influential pioneering work.  String theory remains as a potentially viable candidate in offering a quantum theory of gravity, but is accompanied by the stringent requirement of residing in a higher-dimensional spacetime.  The complete lack of observational evidence of these hypothesized extra dimensions has continued to remain a mystery.  Traditionally, these extra dimensions have been hypothesized to be compactified objects, as is the extra dimension of Kaluza-Klein theory, and hence possibly elude observation due to having an incredibly tiny size.  In the late 1980s, Dai et al. and Hor\v{a}va offered an alternative explanation for the hidden extra dimensions with the discovery of D(irichlet)-branes as fundamental extended objects, with the idea being that our universe may be a D3-brane residing in a higher-dimensional spacetime.\cite{Polchinski}   This discovery has lead to several different braneworld scenarios with large extra spatial dimensions.\cite{Large} 

Over the last few decades, cosmology has evolved into more of a precision science as recent observations have continued to test the predictions of theoretical cosmology.  Observations of distant Type Ia supernovae, precise measurements of the size and spectral distribution of the cosmic microwave background, and the accurate mapping of galaxies in the night sky collectively indicate that the universe, on the largest of scales, is flat, homogenous and isotropic, and is currently undergoing an epoch of accelerated expansion.\cite{Accel,CMB,SDSS} These observations, when compared with standard 4D Friedmann-Robertson-Walker (FRW) cosmology, have constrained the six independent parameters of the Lambda-CDM model and, consequently, have upheld the model as a consistent cosmological theory.  The current accelerated expansion of the universe can be understood through hypothesizing the existence of a vacuum energy density, which must comprise $\sim 70$\% of the total energy and matter content of the universe, as determined through this parameter fitting.  According to standard FRW cosmology, a non-zero vacuum energy would dominate over pressureless matter and radiation in the late universe and give rise to an accelerated expansion via a negative pressure.  Vacuum energy is predicted in quantum field theory as resulting through the continuous creation and annihilation of particle-antiparticle pairs.  However, the ratio of the theoretically estimated value to that which is determined via observation and parameter fitting equates to a factor of $\sim 10^{120}$; this ratio remains as the second worst discrepancy between theory and experiment in history, after the ultraviolet catastrophe of the blackbody spectrum of the late 19th/early 20th century.\cite{CosCon}   

Although remarkably successful at yielding a consistent cosmological picture of the complete time evolution of our universe, standard FRW cosmology leaves several mysteries associated with physical observations unanswered.  
Although a flat, homogeneous and isotropic universe is certainly allowed within the framework of general relativity, these features force a seemingly unnatural fine tuning of the initial conditions of the hot big bang model in the very early universe.  Alternatively, a hypothesized epoch of early-time accelerated expansion, or inflation, neatly explains these phenomena, without the need for fine tuning.\cite{Inflation}  Scalar-field inflationary theory, which generates the desired early-time inflationary epoch, equates as a supplement to standard 4D FRW cosmology, where the energy and matter content of the early universe is assumed to be dominated by a homogenous scalar field over the fluid contributions of radiation, matter, and vacuum energy.  This hypothesized scalar field gives rise to a negative pressure, which drives an early-time inflationary epoch, when the kinetic energy of the scalar field is dominated by its associated potential energy.  This inflationary epoch occurs when the scalar-field `slowly rolls' down its potential energy curve and abruptly ends when the slow roll conditions cease to be met, leading to an era of reheating and particle production.  Although highly successful at solving the horizon and flatness problem, and at giving an explanation for the origin of structure in the universe, the theory of inflation is incomplete as the potential energy function is left undetermined, although it should be noted that these features exist for a wide range of potentials.  Time scales describing the beginning and end of inflation, each respectively defined by the domination of the potential or kinetic energy of the scalar field, are hence left unpredicted by the theory but are hypothesized to equate to when the universe was approximately $10^{-36}$ to $10^{-34}$ seconds old.

Shortly after the advent of scalar-field inflationary theory, higher-dimensional cosmological models were studied as to whether they could potentially account for the large observed entropy of the universe.\cite{others0}  These models were shown to predict inflation of the 3D space during a radiation-dominated epoch of the early universe, hence possibly offering an alternative explanation for inflation to that of scalar-field inflationary theory.  
There, they considered a higher-dimensional spacetime described by a Robertson-Walker metric, where the positively-curved, higher-dimensional compact spatial manifold was allowed to evolve at a different rate than that of the flat 3D noncompact space.
There it was shown that when the higher-dimensional compact space undergoes an era of recollapse towards a (suspected) minimum value, the 3D scale factor rapidly expands.  Since the publication of these original papers, several additional articles have addressed inflation within higher-dimensional spacetime scenarios, each subjected to varying assumptions and/or approximations.\cite{Levin,others3} 

In 2002, Mohammedi offered a possible higher-dimensional alternative explanation to that of vacuum energy driving a late-time accelerated expansion via dynamical compactification, where the higher-dimensional scale factor was assumed to evolve as the inverse of a power of the 3D scale factor.\cite{Mohammedi} 
There, the metric described in the aforementioned paragraph is subjected to a perfect fluid stress-energy tensor, where the pressure in the higher-dimensional space is allowed to differ from that of the 3D space.  The $D$-dimensional FRW field equations were shown to reduce precisely to a 4D form once an effective pressure is defined. The solution for the 3D scale factor was found to undergo accelerated expansion in the late universe for positive 3D and higher-dimensional pressures under this assumption of dynamical compactification, where the higher-dimensional scale factor evolves through an inverse power law. 
Additional papers on anisotropic evolution have emerged in recent years, some including extensions of this dynamical compactification scenario. \cite{Middleton,Pan,others,others2} Other papers have discussed observational constraints placed on large extra dimensions via the Large Hadron Collider and multi-messenger gravitational wave events.\cite{others4}

What this manuscript offers beyond that already found in the literature begins with the discovery of the exact solution to the $D$-dimensional Einstein field equations subjected to a flat, anisotropic Robertson-Walker metric, where the 3D and higher-dimensional scale factors are allowed to evolve at different rates.  Here, the energy-matter content of the $D$-dimensional universe is treated as a perfect fluid, where the pressures in the 3D and higher-dimensional spaces are allowed to differ in the general case.  By adopting two equations of state linearly relating the 3D and higher-dimensional pressures to the density, the $D$-dimensional FRW field equations are decoupled and an exact expression relating the higher-dimensional scale factor to a function of the 3D scale factor is obtained. This decoupling allows for the construction of the effective 4D FRW field equations, written solely in terms of the 3D scale factor, which ultimately can be solved exactly for a single fluid component.  This exact treatment allows for the identification of two limiting regimes, which ultimately describes the behavior of the early- and late-time universe.

The fluid regime solution is found to closely mimic that of 4D FRW cosmology, offering a late-time behavior for the 3D scale factor after becoming valid in the early universe, and can give rise to a late-time accelerated expansion driven by vacuum energy.  Interestingly, the lowest-order contribution of the series solution of this regime equates to a generalization of the dynamical compactification scenario of Mohammedi \cite{Mohammedi}.  Here it is shown that the higher-dimensional scale factor evolves as the inverse of a power of the 3D scale factor for a limited range of 3D and higher-dimensional equation of state parameters.  We further show that accelerated expansion of the 3D scale factor is only obtained for a limited range of \textit{negative} 3D and higher-dimensional equation of state parameters in this generalized treatment and that if the 3D scale factor is in fact undergoing accelerated expansion in this regime, then the higher-dimensional scale factor must be simultaneously expanding.

The volume regime yields two branches of solutions, where two distinct cases arise for each branch.  These correspond collectively to a generalization of the $D$-dimensional accelerating vacuum solutions of Levin and decelerating vacuum solutions of Chodos and Detweiler, where the time dependence of the 3D and higher-dimensional scale factors are merely functions of the number of spacetime dimensions.\cite{Levin,Chodos}  In this manuscript, we show that both cases of each branch of solutions can in fact arise, but only for a limited range of 3D and higher-dimensional EoS parameters, $w$ and $v$.  For the unique case of $d=1$, the volume regime solution is shown to offer a very early-time epoch of accelerated expansion, when radiation is the dominant energy component, for any negative value of the higher-dimensional equation of state parameter.  This volume regime solution is shown to turn on and then off, hence offering a natural entrance and exit from a possible inflationary epoch.  

The fluid regime solution is then shown to turn on only after the volume regime solution turns off and consequently remains on indefinitely. 
This time ordering is obeyed so long as the predicted constant initial 3D scale factor of the volume regime is larger than a threshold value, where this lower bound is a function of the higher-dimensional equation of state parameter.   Further, we show that the time scales marking the end of the volume regime and the beginning of the fluid regime are constrained to reside within a predicted range, with both times scales bounded from below and in some cases from above.  By aligning the fundamental time constant of the theory with the Planck time and the corresponding constant initial 3D scale factor with the value predicted by standard 4D FRW cosmology at this time, we show that the aforementioned time scales marking the end of the volume regime and the beginning of the fluid regime take on maximum values of $t_{\scriptsize{\mbox{vol},f,\mbox{max}}}\sim 10^{-35.5}\mbox{s}$ and $t_{\scriptsize{\mbox{flu},i,\mbox{max}}}\sim 10^{-13}\mbox{s}$.  Curiously, these time scales match remarkably well with the times predicted for the unification of the strong and electroweak force and the unification of the electromagnetic and weak force, respectively.

This paper is organized as follows.  In Sec. \ref{framesec}, we subject the $D=d+4$ dimensional Einstein field equations to a flat, anisotropic Robertson-Walker metric.  By adopting two equations of state, we arrive at an expression for the higher-dimensional scale factor as a function of the 3D scale factor.  In Sec. \ref{eff}, we write the $D$-dimensional FRW field equations exclusively in terms of the 3D scale factor, arriving at a set of effective 4D field equations.  These equations can be solved exactly, however, we withhold the presentation of the exact treatment until Appendix \ref{GT}.  In Sec. \ref{fluid}, we present the fluid regime solution, which equates to a generalized treatment of the dynamical compactification scenario of Mohammedi\cite{Mohammedi}.  In Sec. \ref{vacuum1}, we present the volume regime solutions, which equate to a generalized treatment of the $D$-dimensional accelerating vacuum solutions of Levin and decelerating vacuum solutions of Chodos and Detweiler.\cite{Levin,Chodos} In Sec. \ref{inflation}, we study the strong inequalities that define the fluid and volume regimes for $d=1$.  We show that one of the two volume regime solutions gives rise to an epoch of accelerated expansion in the very early universe for the 5D case and offers a possible higher-dimensional alternative to scalar-field inflationary theory.  Finally, in Sec. \ref{conclusion} we summarize our results.

\section{Anisotropic $D$-dimensional FRW cosmology}\label{framesec}

We begin with the Einstein field equations in $D=d+4$ spacetime dimensions of the form 
\beq\label{field}
 G_A{}^B=\frac{8\pi G_{\tiny{D}}}{c^2}T_A{}^B,
  \eeq
where $A,B$ are indices that run over all spacetime dimensions and $G_{\tiny{D}}$ is the higher-dimensional Newtonian constant.  For notational simplicity, we set this coupling constant $8\pi G_{\scriptsize{D}}/c^2$ equal to one.   The higher-dimensional stress-energy tensor will be assumed to be that of a perfect fluid and of the form
\beq\label{stress}
T_A {}^ B= \textrm{diag} \left[ -\rho(t), p(t) , p(t), p(t), p_d(t), ... ,p_d(t)\right],
\eeq
where $p(t)$ and $p_d(t)$ are the pressures of the 3D and higher-dimensional spaces, respectively.  As is obvious from Eq. (\ref{stress}), we are allowing the pressure in the higher-dimensional space to be different, in general, from the pressure in the 3D space.  Hence, this stress-energy tensor describes an anisotropic perfect fluid in $d+4$ spacetime dimensions.

We choose a metric ansatz of the form
\bea\label{metric} 
ds^2&=&- dt^2+a^2(t)
\left(dr^2 + r^2 \left(d \theta^2 + \sin^2\theta d \phi^2 \right) \right)\nonumber\\
&&+b^2(t)(dy_1^2+dy_2^2+...+dy_d^2), 
\eea
where we allow the scale factor of the higher-dimensional manifold, $b(t)$, to evolve at a different rate, in general, than the 3D scale factor, $a(t)$.   This metric ansatz describes flat, homogeneous and isotropic 3D and higher-dimensional spaces, where we work in units where the speed of light is set equal to unity.

By adopting the above perfect fluid stress-energy tensor and metric ansatz, the $D$-dimensional FRW field equations and the $D$-dimensional conservation equation are of the form
\bea
\rho&=&3\frac{\dot a^2}{a^2} +\frac{1}{2}d(d-1)\frac{\dot b^2}{b^2}+3d\frac{\dot a\dot b}{ab}\label{density}\\
p&=&- \left[2 \frac{\ddot a}{a} +\frac{\dot a^2}{a^2}+d\frac{\ddot b}{b}+\frac{1}{2}d(d-1)\frac{\dot b^2}{b^2}+2d\frac{\dot a\dot b}{ab}\right] \label{pres}  \\
p_d&=&- \left[3\left( \frac{\ddot a}{a} +\frac{\dot a^2}{a^2}\right)+(d-1)\left(\frac{\ddot b}{b}+\frac{1}{2}(d-2)\frac{\dot b^2}{b^2}+3\frac{\dot a\dot b}{ab}\right)\right]\;\;\;\;\label{presD}\\
0&=&\dot{\rho}+3\frac{\dot{a}}{a}(\rho+p)+d\frac{\dot{b}}{b}(\rho+p_d),\label{cons}
\eea
where a dot denotes a time derivative.  One can easily show that the conservation equation, given by Eq. (\ref{cons}), is in fact satisfied when Eqs. (\ref{density})-(\ref{presD}) are employed.  Hence, the physical variables of the field equations, for general density and pressures, are left underdetermined as we have three unique equations and five unknowns. This is analogous to standard 4D FRW cosmology where one typically adopts an equation of state (EoS) connecting the pressure to the density, hence generating a determinable system of equations.  Here, we follow a similar treatment by adopting two equations of state of the form
\bea
p&=&w\;\rho\label{EoS}\nonumber\\
p_d&=&v\;\rho,\label{EoSd}
\eea
linearly relating the 3D and higher-dimensional pressures to the density.  Notice that the 3D and higher-dimensional EoS parameters, $w$ and $v$, can in general be time-dependent, in this manuscript they are simply treated as constants.  Also notice that here we limit the EoS parameters to the domain $-1\leq w,v<1$. Lastly, notice that Eqs. (\ref{stress}), (\ref{metric}) and (\ref{EoS}) equate to the main assumptions of this manuscript.

Remarkably, the scale factors $a(t)$ and $b(t)$ can be decoupled from one another.  Using Eq. (\ref{EoS}) to eliminate $\rho$, $p$, and $p_d$ from Eqs. (\ref{density})-(\ref{presD}) and performing some algebra, we obtain an exact differential equation of the form
\bea\label{diff}
\frac{d}{dt}\left[a^{3-dn}\;\frac{d}{dt}\left(a^nb\right)^d\right]&=&0,
\eea

\begin{figure}[!h]
\begin{center}
\includegraphics[width=12cm]{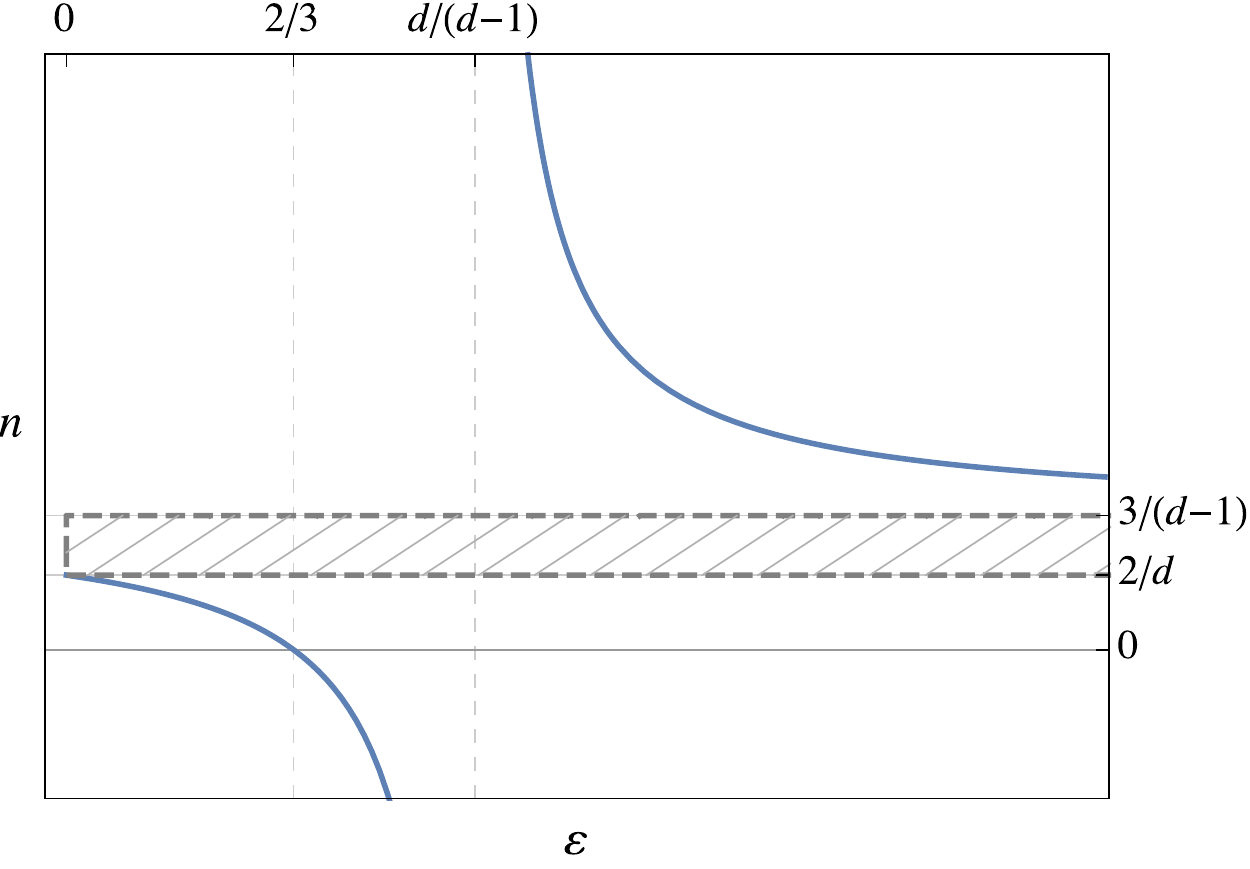}
\caption{Plot of $n$ versus $\varepsilon$ for $d>1$.  Notice that $n$ takes on positive values in two distinct regions (where $\varepsilon<2/3$ and $\varepsilon>d/(d-1)$), negative values in the region where $2/3<\varepsilon<d/(d-1)$, and diverges as $\varepsilon\rightarrow d/(d-1)$.  Also notice that $n$ \textit{only} takes on values given by $n\leq 2/d$ or $n> 3/(d-1)$ for $0<\varepsilon<\infty$.  The hashed region indicates the forbidden range of values for $n$ as $\varepsilon>0$.  This plot was generated for $d=6$, however, the same basic features persist for all $d>1$.}
\label{fig:nvseps}
\end{center}
\end{figure}

where we defined the power
\beq\label{n} 
n\equiv\left[\frac{2-3\varepsilon}{d-(d-1)\varepsilon}\right]\;\;\mbox{with}\;\;\varepsilon\equiv\frac{(1-w)}{(1-v)}.
\eeq
Notice that $\varepsilon$ lies within the range $0<\varepsilon<\infty$ and hence is strictly positive as here we consider EoS parameters within the domain $-1\leq w,v< 1$.  Also notice that when the 3D and higher-dimensional EoS parameters, $w$ and $v$, are set equal to one another, the aforementioned power takes on the value $n=-1$.  See Fig. \ref{fig:nvseps} for a plot of the power $n$ versus $\varepsilon$, where both are defined in Eq. (\ref{n}).  Notice that $n$ takes on positive values in two distinct regions specified by $\varepsilon<2/3$ and $\varepsilon>d/(d-1)$ and negative values in the region defined by $2/3<\varepsilon<d/(d-1)$ for all $d>1$.  Also notice that $n$ diverges as $\varepsilon\rightarrow d/(d-1)$, approaching either negative or positive infinity as $\varepsilon$ approaches $d/(d-1)$ from the left or right, respectively.  Interestingly, it is noted that $n$ \textit{only} takes on values $n\leq2/d$ or $n>3/(d-1)$ for our allowed EoS parameter space. Lastly, for the special case of $d=1$, $n$ simplifies to a linear expression in $\varepsilon$ as the denominator of Eq. (\ref{n}) reduces to one and, consequently, $n$ remains positive \textit{only} for $\varepsilon<2/3$.  This special case of $d=1$ was studied by one of the authors in a previous work.\cite{Middleton}

Equation (\ref{diff}) can be integrated twice where the higher-dimensional scale factor can be written as a function of 3D scale factor.  Performing the aforementioned integrations, one finds an expression for the higher-dimensional scale factor of the form
\beq\label{bfuna}
b(t)= \frac{1}{a^{n}(t)}\left[\gamma_1+\gamma_0 \int a(t)^{(dn-3)}dt\right]^{1/d},
\eeq
where $\gamma_0$ and $\gamma_1$ are constants of integration and $n$ is defined in Eq. (\ref{n}).  Notice that although $\gamma_1$ is a unitless numerical scale, $\gamma_0$ has units of inverse seconds.

It should be emphasized that Eq. (\ref{bfuna}) is an \textit{exact} expression for one fluid component, relating the higher-dimensional scale factor to a function of the 3D scale factor.  As is obvious from Eq. (\ref{bfuna}), the higher-dimensional scale factor is determined by two competing terms, each characterized by a corresponding integration constant.  Notice that if $\gamma_0$ is set equal to zero, we obtain a generalization of the dynamical compactification scenario of Mohammedi where, there, $n$ was an arbitrary power that was held positive to ensure contraction of the higher-dimensional scale factor.\cite{Mohammedi}  Here, dynamical compactification emerges naturally as a special case of an exact treatment for a limited range of 3D and higher-dimensional EoS parameters as the value of $n$ is determined by $w$ and $v$ via Eq. (\ref{n}).  

The discovery of Eq. (\ref{bfuna}) allows us to decouple the $D$-dimensional FRW equations and obtain a set of differential equations written exclusively in terms of the 3D scale factor.  In the next section, we first present this set of effective 4D FRW field equations prior to obtaining the exact solution for the 3D scale factor and, more illuminating, the approximate solutions in two distinct regimes. 

\section{Effective 4D FRW field equations}\label{eff}
Now that we have an expression for the higher-dimensional scale factor in terms of the 3D scale factor, we can write the $D$-dimensional FRW field equations solely in terms of $a(t)$.  After performing some algebra, we find that Eqs. (\ref{density})-(\ref{cons}) can be written in the form
\bea\label{4Deff}
\rho&=&\eta_1\frac{\dot a^2}{a^2}-\frac{1}{dn}(2\eta_1+3\eta_2)\frac{\gamma_0}{x}\frac{\dot a}{a}+\frac{1}{2d}(d-1)\frac{\gamma_0^2}{x^2}\label{rho0}\\
\tilde{p}&=&-\frac{1}{3}\eta_1\left(2\frac{\ddot a}{a}+\frac{\dot a^2}{a^2}\right)\nonumber\\
&&-\left[\frac{1}{dn}(2\eta_1+3\eta_2)-\frac{2}{3}(\eta_1+3\eta_2)\right]\frac{\gamma_0}{x}\frac{\dot a}{a}\nonumber\\
&&-\frac{1}{3}(1+2\eta_2)\cdot\frac{1}{2d}(d-1)\frac{\gamma_0^2}{x^2}\label{p0}\\
p_d&=&\frac{1}{dn}\left[(2\eta_1+3\eta_2)\frac{\ddot a}{a}+(2\eta_1-\eta_2(\eta_1+3\eta_2))\frac{\dot a^2}{a^2}\right]\nonumber\\
&&+\frac{1}{2d}(d-1)\frac{\gamma_0^2}{x^2}\label{pd0}\\
0&=&\dot{\rho}+3(1+\tilde{w})\frac{\dot{a}}{a}\rho+(1+v)\frac{\gamma_0}{x}\rho,\label{cons0}
\eea
where we defined the parameters
\bea
\eta_1&\equiv&\frac{1}{2}d(d-1)(n-\alpha_+)(n-\alpha_-)\label{eta1}\\
\eta_2&\equiv&(dn-2)\label{eta2},
\eea
where we used Eq. (\ref{n}) and defined the roots of $\eta_1$ as
\beq\label{alphapm}
\alpha_\pm\equiv\frac{1}{(d-1)}\left[3\pm\sqrt{\frac{3(d+2)}{d}}\;\right].
\eeq
It is noted that these roots are positive for all $d$. As is evident from Eq. (\ref{eta1}), the parameter $\eta_1$ takes on positive values in two distinct regions, where $n>\alpha_+$ and $n<\alpha_-$, and negative values in the region where $\alpha_-<n<\alpha_+$.   
Notice that for the special case of $d=1$, $\eta_1$ becomes linear in $n$ as the $\alpha_+$ root diverges whereas $\alpha_-$ becomes unity.  For $d>1$, the $\alpha_\pm$ roots take on a limited range of values constrained by $3/(d-1)<\alpha_+$ and $0<\alpha_-<2/d$. 


Notice that Eq. (\ref{p0}) represents an effective pressure and equates to a linear combination of Eqs. (\ref{density})-(\ref{presD}), which can be shown with the help of Eq. (\ref{bfuna}).  This effective pressure presented in Eq. (\ref{p0}) is generated through a combination of the density and pressures given by
\beq\label{effective}
\tilde{p}\equiv p-\frac{1}{3}dn(\rho+p_d)=\tilde{w}\rho,
\eeq
where $\tilde{w}$ represents an effective EoS parameter.  By defining this effective pressure, notice that the $D$-dimensional conservation equation given by Eq. (\ref{cons0}) reduces to a form that precisely mimics the 4D conservation equation with $w$ replaced with $\tilde{w}$ once the integration constant $\gamma_0$ is set equal to zero or for the special case of $v=-1$.  This effective pressure was originally defined elsewhere, but proves useful in our analysis of the following sections.\cite{Mohammedi}  Also notice that if one redefines the coupling constant (which was previously set equal to one for notational simplicity) to absorb the parameter $\eta_1$, Eqs. (\ref{rho0}) and (\ref{p0}) also mimic that of 4D FRW cosmology with $p$ replaced with $\tilde{p}$, when $\gamma_0$ is set equal to zero.

In arriving at the effective 4D FRW field equations of Eqs. (\ref{rho0})-(\ref{pd0}), we defined the higher-dimensional volume element, $x$, as 
\beq\label{volume}
 x(a)\equiv a^3b^d=a(t)^{(3-dn)}\left[\gamma_1+\gamma_0\int a(t)^{(dn-3)}dt\right],
\eeq
where we again used Eq. (\ref{bfuna}) to express this volume element solely in terms of 3D scale factor, $a(t)$.  Hence, the effective 4D FRW field equations are non-linear ODEs that contain integrals of the scale factor via the $\gamma_0/x$ terms (and squares of the integral of the scale factor via the $\gamma_0^2/x^2$ terms for $d\neq 1$) and therefore effectively equate to third-order differential equations for the integral. It is interesting to note that the $\gamma_0^2/x^2$ terms are somewhat analogous to the curvature terms of standard non-flat 4D FRW cosmology.  These volume terms compete for dominance with the density and pressure terms as they collectively determine the time evolution of the scale factor.  It is further noted that these curvature-like $\gamma_0^2/x^2$ terms are only present for $d>1$, which add to the intricacy of the field equations.

As a check of the effective 4D field equations, one can solve Eqs. (\ref{rho0}) and (\ref{p0}) for $\dot a^2/a^2$ and $\ddot a/a$ in terms of the density, pressure and the higher-dimensional volume element and then substitute these expressions into Eq. (\ref{pd0}).  Using Eq. (\ref{EoS}), the parameter relations 
\bea\label{parametersw}
3(1+\tilde{w})(1+\alpha)=\frac{2\eta_1}{\eta_2}(1-w)=\frac{2\eta_1(1-v)}{((d-1)n-3)}&&\nonumber\\
\mbox{where}\;\;\alpha\equiv\frac{2(dn-3)}{3(1+\tilde{w})},&&
\eea
and performing some algebra, one can in fact verify that Eq. (\ref{pd0}) is a redundant expression, as should be expected.

Similar to 4D FRW cosmology, the $D$-dimensional conservation equation yields an expression for the density as a function of the 3D scale factor.  Integrating Eq. (\ref{cons}) and using Eq. (\ref{bfuna}), we arrive at an expression for the higher-dimensional density of the form
\beq\label{rhofuna}
\rho(a)=\frac{\rho_0}{a(t)^{3(1+\tilde{w})}}\left[\gamma_1+\gamma_0\int a(t)^{(dn-3)}dt\right]^{-(1+v)},
\eeq
where $\rho_0$ is another constant of integration.  Notice that in order to obtain a constant energy density for our allowed domain of EoS parameters, we must set $w=v=-1$; this becomes evident upon close examination of Eqs. (\ref{n}), (\ref{effective}) and (\ref{rhofuna}).  Also notice that when the higher-dimensional EoS parameter is set to $v=-1$, Eqs. (\ref{cons}) and (\ref{cons0}), and consequently Eq. (\ref{rhofuna}), reduce precisely to the 4D form.  This special case equates to the first law of thermodynamics for cosmology in 4D.  Hence, for the special case of $v=-1$, the dynamical evolution of the density is completely independent of the extra dimensions.


Before proceeding to a general treatment of the effective 4D field equations, we first discuss the conditions necessary for accelerated expansion.  Combining Eqs. (\ref{rho0}) and (\ref{p0}), we arrive at an expression for the acceleration of the 3D scale factor of the form
\bea\label{acceleration}
\frac{\ddot a}{a}&=&-\frac{1}{2\eta_1}(1+3\tilde{w})\rho+\left[1+\frac{(d+2)n}{\eta_1}\right]\frac{\gamma_0}{x}\frac{\dot a}{a}\nonumber\\
&&-\frac{(d-1)}{2d}\frac{\eta_2}{\eta_1}\frac{\gamma_0^2}{x^2}.
\eea
As is obvious from Eq. (\ref{acceleration}), the value of this acceleration is determined by three competing terms.  When the first term on the right-hand side dominates, one obtains accelerated expansion for $-1\leq\tilde{w}<-1/3,\;\eta_1>0$, and a positive energy density.
In the next section, we present the series solution for the 3D scale factor in this aptly named \textit{fluid regime}, which yields this accelerated expansion scenario.  We show that accelerated expansion in this regime is only obtained for a limited range of \textit{negative} 3D and higher-dimensional EoS parameters, which corresponds to higher-dimensional dark energy or vacuum energy when $w=v=-1$.
We also show that if the 3D scale factor is in fact undergoing accelerated expansion in this regime, then the higher-dimensional scale factor must be simultaneously expanding.  

For the unique case of $d=1$, the third term on the right-hand side of Eq. (\ref{acceleration}) vanishes and the value of the acceleration is consequently determined by the two remaining terms.  Interestingly, we note that the second term on the right-hand side is positive, yielding accelerated expansion if dominant, for $\eta_1>0$, $\gamma_0>0$, and an expanding 3D space.  In Sec. \ref{vacuum1}, we find the series solution for the 3D scale factor in this \textit{volume regime}, which yields this aforementioned scenario of second-term domination and accelerated expansion when $d=1$.  This corresponding solution equates to one of two branches of solutions and will be referred to as the $\alpha_+$ solution, where $\gamma_0>0$ for all allowed values of the EoS parameters.  

Astonishingly, the effective 4D field equations of Eqs. (\ref{rho0})-(\ref{pd0}) can be solved \textit{exactly} for a single fluid component, meaning they can be integrated thrice for an integral of the 3D scale factor raised to a power.  The resultant expression relates a function of the 3D scale factor, namely a product of a power of an integral of the 3D scale factor and a hypergeometric function, to the time and proves to be rather cumbersome in a general analytical treatment.  As hypergeometric functions contain singularities, and these singularities can be expanded about by writing in terms of a hypergeometric series, the exact solution reduces to more convenient series solutions in ultimately two distinct regimes.  This exact treatment is withheld until Appendix \ref{GT}.

Equivalently, one can solve the effective 4D FRW field equations perturbatively in each of these two distinct  
regimes, each characterized by a set of strong inequalities that ultimately define their relevant time regimes.  In the following sections, we present the series solutions for the 3D and higher-dimensional scale factors in the fluid and volume regimes.  We show that the fluid regime solution closely mimics that of standard 4D FRW cosmology, gives rise to a late-time accelerated expansion driven by a higher-dimensional vacuum energy, and can become valid in the early universe following an even earlier epoch of volume regime solution validity.  We then show that the volume regime yields two branches of solutions, where two distinct cases arise for each branch, which collectively correspond to a generalization of the $D$-dimensional accelerating vacuum solutions of Levin and decelerating vacuum solutions of Chodos and Detweiler.\cite{Levin,Chodos}  We show that both cases of each branch of solutions can arise, but only for a limited range of EoS parameters.  For the unique case of $d=1$, we show that one of the two volume regime solutions offers a very early epoch of accelerated expansion for the 3D scale factor for \textit{any} 3D EoS parameter value, $w$, as long as $v<0$.  This volume regime solution is valid in an early-time radiation-dominated epoch and gives rise to a potential higher-dimensional alternative to scalar-field inflation, as the solution is shown to naturally turn on and off, hence offering a graceful exit from an inflationary epoch.  This solution is followed by the relatively late-time fluid regime solution.  The time scales for this transition are shown to be within a small fraction of the first second when the fundamental constants of the theory are aligned with the Planck scale, with the bounds determined by the values of the constant initial 3D scale factor and the higher-dimensional EoS parameter. 

\section{The fluid regime}\label{fluid}

Here we are interested in the approximate solution to the effective 4D FRW field equations of Eqs.  (\ref{rho0})-(\ref{pd0}) for the 3D and higher-dimensional scale factors in the regime when
\bea\label{fluidregime}
\rho&\gg& \frac{\gamma_0}{x}\frac{\dot{a}}{a}\gg\frac{\gamma_0^2}{x^2}\;\;\mbox{as}\;\;\frac{\dot{a}}{a}\gg\frac{\gamma_0}{x}\;\;\mbox{and}\nonumber\\
&&\gamma_1\gg\gamma_0\int a(t)^{(dn-3)}dt.
\eea
In this regime, the density and pressure components dominate over terms involving the inverse of the higher-dimensional volume element.  The effective field equations reduce to a form similar to that of standard 4D FRW cosmology at lowest order.  Here, the time evolution of the 3D scale factor closely mimics that predicted by standard 4D cosmology and the solution in this regime is shown to become valid in the early universe and remain so indefinitely.  
It should be noted that the above strong inequalities are in reference to the magnitude of the terms only; at this point the constants of integration $\gamma_0,\gamma_1$ can be of either sign.  

In this regime, the higher-dimensional scale factor of Eq. (\ref{bfuna}) takes on the approximate form 
\beq\label{moham}
b(t)\sim\frac{1}{a^n(t)}
\eeq
at lowest order.  Hence, this regime generalizes the dynamical compactification scenario first investigated by Mohammedi\cite{Mohammedi}, where here the value of $n$ is determined by the 3D and higher-dimensional EoS parameters as demanded by Eq. (\ref{n}).  The effective 4D FRW field equations can be solved perturbatively by writing the 3D scale factor as a series solution of the form
\beq\label{perturb1}
a(t)=a_0(t)+\left(\frac{\gamma_0}{\gamma_1}\right)a_1(t)+...
\eeq
where $a_0(t)$ is the zeroth-order solution, $a_1(t)$ is the first-order correction term, etc.  Notice that in this regime, the $\gamma_0$ terms in the effective field equations of Eqs. (\ref{rho0})-(\ref{pd0}) do not yield a contribution to the zeroth-order solution as these $\gamma_0$ terms contribute to the first-order solution at lowest order.  Similarly, the $\gamma_0^2$ terms in Eqs. (\ref{rho0})-(\ref{pd0}) first yield a contribution to the second-order solution at lowest order.  

Inserting Eq. (\ref{perturb1}) into Eqs. (\ref{rho0})-(\ref{pd0}), grouping by order of $\gamma_0/\gamma_1$ and solving, we obtain the desired series solution for the 3D scale factor, which takes the form
\bea\label{asol0}
&&a(t)=\tilde{a}_0t^{2/3(1+\tilde w)}\hspace{2.6cm}\mbox{for}\;\;\tilde{w}\neq -1\nonumber\\
&\cdot&\left[1-\left(\frac{\gamma_0}{\gamma_1}\right)\frac{2v\cdot \tilde{a}_0^{dn-3}}{3(1+\tilde w)(1+\alpha)(2+\alpha)}t^{(1+\alpha)}+...\right]
\eea
where the integration constant $\tilde{a}_0>0$ and should not be confused with the zeroth-order perturbative solution, $a_0(t)$, as defined in Eq. (\ref{perturb1}).  Notice that $\tilde{a}_0$ can be fixed by normalization so that $a(t_0)=1$ when evaluated at the present time.  Also notice that to lowest order, this solution for the 3D scale factor has the same functional form as that of standard 4D FRW cosmology with $w$ replaced with the effective EoS parameter, $\tilde{w}$, which was previously defined in Eq. (\ref{effective}).   This effective EoS parameter can potentially lie within the range $-1\leq\tilde{w}<-1/3$, hence yielding accelerated expansion for the 3D scale factor, for positive values of $w$ and $v$.  We will explore this possibility later in this section.  

For Eq. (\ref{asol0}) to be a valid and consistent solution to all three of the effective 4D FRW field equations, the integration constant $\tilde{a}_0$ is found to obey the algebraic relation
\beq\label{a0rho0}
\rho_0\gamma_1^{-(1+v)}=\left[\frac{2}{3(1+\tilde{w})}\right]^2\eta_1\tilde{a}_0^{3(1+\tilde{w})}\;\;\mbox{when}\;\;\tilde{w}\neq -1.
\eeq
Notice that Eq. (\ref{a0rho0}) agrees with the relation one obtains from standard 4D FRW cosmology with the exception of the additional factors $\gamma_1$ and $\eta_1$.  

Now, inserting Eq. (\ref{asol0}) into Eq. (\ref{bfuna}), the higher-
dimensional scale factor takes the form
\bea\label{bfunasol0}
&&b(t)=\tilde{b}_0 t^{-2n/3(1+\tilde{w})}\hspace{3.5cm}\mbox{for}\;\;\tilde{w}\neq -1\nonumber\\
&&\cdot\left[1+\left(\frac{\gamma_0}{\gamma_1}\right)\frac{\tilde{a}_0^{(dn-3)}}{d(1+\alpha)}\left[1+\frac{2v\cdot dn}{3(1+\tilde{w})(2+\alpha)}\right]t^{(1+\alpha)}+...\right],\nonumber\\
\eea
where the coefficient of the higher-dimensional scale factor in this fluid regime is determined by
\beq\label{tildeb0}
\tilde{b}_0\equiv\frac{\gamma_1^{1/d}}{\tilde{a}_0^n}.
\eeq
Notice that if one demands that the higher-dimensional scale factor be positive and real, then we arrive at the requirement that $\gamma_1>0$.  Also notice that $\gamma_1$ plays the interesting role of an overall numerical factor that scales the time-dependent expression for the higher-dimensional scale factor given by Eq. (\ref{bfunasol0}).  Although the sign of $n$ determines whether the higher-dimensional scale factor expands or contracts at lowest order, a vanishingly small value of $\gamma_1$ can scale away its overall significance in the line element, given by Eq. (\ref{metric}), and possibly offers an explanation for its absence from observation.

\subsection{Positive energy density}

The requirement that the higher-dimensional energy density be positive is equivalent to the parameter constraint $\eta_1>0$.  This can be easily witnessed through Eq. (\ref{a0rho0}) as both $\gamma_1$ and $\tilde{a}_0$ are positive as we require the 3D and higher-dimensional scale factors to be both positive and real.  Employing  Eqs. (\ref{n}) and (\ref{eta1}), we find two possible cases of EoS parameter inequalities given by 
\bea
v&<&1-\frac{(d-1)}{d}\left[\frac{\alpha_+-3/(d-1)}{\alpha_+-2/d}\right](1-w)\nonumber\\
&&\mbox{and}\;\;v>1-\frac{(d-1)}{d}(1-w)\;\;\;\;\;\;\;\mbox{or}\label{inequal1}\\
v&>&1-\frac{(d-1)}{d}\left[\frac{\alpha_--3/(d-1)}{\alpha_--2/d}\right](1-w)\nonumber\\
&&\mbox{and}\;\;v<1-\frac{(d-1)}{d}(1-w),\label{inequal1also}
\eea
where the parameters $\alpha_\pm$ were defined in Eq. (\ref{alphapm}).  Notice that these two distinct cases arise from the fact that $\eta_1>0$ if $n<\alpha_-$ or if $n>\alpha_+$.  Special care must be taken in arriving at Eqs. (\ref{inequal1}) and (\ref{inequal1also}) as the denominator of $n$ can be positive or negative, which can be easily seen through examination of Eq. (\ref{n}).  

Figure \ref{fig:all} shows a plot of the allowed parameter space subjected to the two cases of dual inequalities of Eqs. (\ref{inequal1}) and (\ref{inequal1also}).  The hashed regions equate to $\eta_1<0$, hence a negative energy density, and are not considered in this manuscript.  Although this EoS parameter space plot was generated for $d=6$, the same basic features remain for all values of $d>1$.  Notice that for the unique case of $d=1$, the upper wedge-like hashed region vanishes and all values of the EoS parameters in the second quadrant are consequently allowed.  


\subsection{Accelerated expansion of the 3D scale factor}

Accelerated expansion occurs for the perturbative 3D solution analyzed in this section when the effective EoS parameter resides in the range $-1\leq\tilde{w}<-1/3$, which can be seen by examining the power of the zeroth-order solution given in Eq. (\ref{asol0}) or Eq. (\ref{acceleration}).  Employing Eqs. (\ref{n}) and (\ref{effective}), we find accelerated expansion for the 3D scale factor when the set of inequalities
\bea\label{inequal3}
0&\leq&\left(v-\frac{3}{2}w\right)^2-\frac{3(d+2)}{4d}w^2+\frac{(d+3)}{2d},\nonumber\\
0&>&\left(v-\frac{3}{2}\left(w-\frac{1}{3}\right)\right)^2-\frac{3(d+2)}{4d}\left(w-\frac{1}{3}\right)^2\nonumber\\
&&+\frac{(d+2)}{3d}\;\;\;\;\;\mbox{and}\nonumber\\
v&<&1-\frac{(d-1)}{d}(1-w)
\eea
are collectively satisfied.  Figure \ref{fig:all} shows a plot of the EoS parameter space that yields  $-1\leq\tilde{w}<-1/3$.  It is interesting to note that the \textit{entire} parameter space considered in the manuscript obeys the first inequality, which equates to $-1\leq\tilde{w}$. Hence, the 3D scale factor expands for all values of  $w$ and $v$ whereas the higher-dimensional scale factor can either expand or contract, depending on the values of  $w$ and $v$, as discussed in the previous subsection.  

\subsection{Dynamical compactification of the higher-dimensional scale factor}

We now wish to examine the EoS parameter space for regions where dynamical compactification occurs.  Dynamical compactification equates to $n>0$ in the fluid regime, where the higher-dimensional scale factor contracts as the 3D scale factor expands, which can be seen via Eq. (\ref{moham}).  Employing Eq. (\ref{n}), we find two possible cases of EoS parameter inequalities given by the relations
\bea
v&>&\frac{1}{2}(3w-1)\;\;\mbox{and}\;\;v>1-\frac{(d-1)}{d}(1-w)\;\;\mbox{or}\label{inequal2}\\
v&<&\frac{1}{2}(3w-1)\;\;\mbox{and}\;\;v<1-\frac{(d-1)}{d}(1-w).\label{inequal2also}
\eea
These two distinct cases arise from the fact that $n>0$ when the numerator and denominator of $n$ share the same sign.

\begin{figure}[!t]
\begin{center}
\includegraphics[width=10cm]{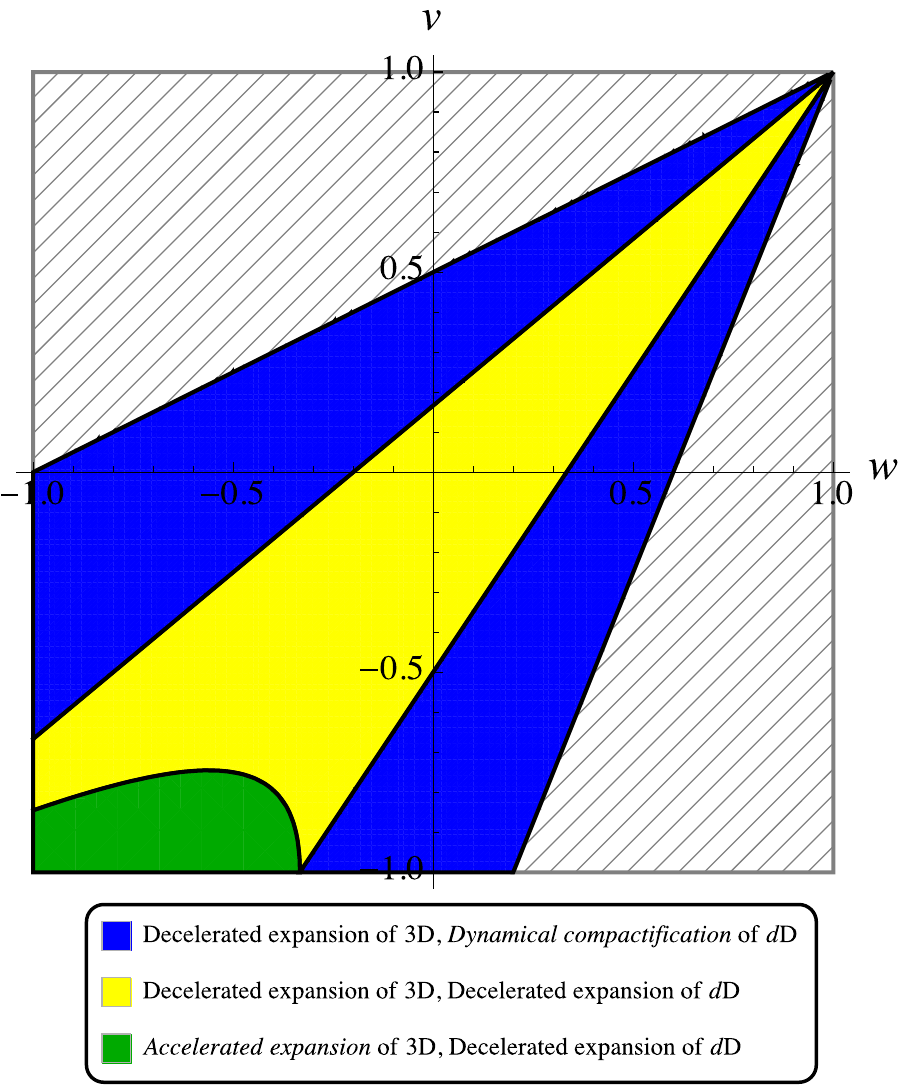}
\caption{EoS parameter space plot of $v$ versus $w$ that indicates the parameter space values that yield accelerated expansion for the 3D scale factor or dynamical compactification of the higher-dimensional scale factor in the fluid regime (or neither).  Notice that no allowed parameter values yield both.   
The hashed regions indicate the values of EoS parameters that yield a negative energy density and are not considered here. This plot was generated for $d=6$, however, the same basic features persist for all $d>1$.}
\label{fig:all}
\end{center}
\end{figure}

Figure \ref{fig:all} shows a plot of the EoS parameter space subjected to both cases of the dual inequalities of Eqs. (\ref{inequal2}) and (\ref{inequal2also}).  
It is interesting to note that the area of the blue lower-right wedge-like region is independent of the number of extra spatial dimensions whereas the area of the blue upper-left wedge-like region varies with $d$.  For the unique case of $d=1$, the blue upper-left wedge-like region vanishes and all values of EoS parameters in the second quadrant of this parameter space give rise to $n<0$.

Finally, we explore the EoS parameter space constrained by the requirement that the energy density remains positive and present the range of EoS parameters that yields dynamical compactification of the higher-dimensional scale factor and accelerated expansion of the 3D scale factor.  
Figure \ref{fig:all} shows a plot of the EoS parameter space.  Notice that neither of the two regions of the EoS parameter space that yield dynamical compactification overlap with the region that yields accelerated expansion of the 3D scale factor.  
Thus, one can conclude that if the 3D scale factor is undergoing accelerated expansion when the perturbative solution of this section is valid, then the higher-dimensional scale factor \textit{cannot} be simultaneously dynamically compactifying.  A similar result was found in the five dimensional case studied elsewhere.\cite{Middleton,Pan}


\section{The volume regime}\label{vacuum1}

Here we are interested in arriving at the approximate solutions to the effective 4D FRW field equations of Eqs. (\ref{rho0})-(\ref{pd0}) for the 3D and higher-dimensional scale factors in the regime when the expressions 
\bea\label{vacuumregime}
\rho&\ll& \frac{\gamma_0}{x}\frac{\dot{a}}{a}\sim\frac{\gamma_0^2}{x^2}\;\;\mbox{as}\;\;\frac{\dot{a}}{a}\sim\frac{\gamma_0}{x}\;\;\mbox{and}\nonumber\\
&&\;\gamma_1\ll\gamma_0\int a(t)^{(dn-3)}dt
\eea
hold, with the unique exception of one of two solutions for $d=1$.  This special case in the volume regime for $d=1$ obeys a modified set of strong inequalities defining its regime of validity and will be further discussed later in this section.  It is important to note that the above strong inequalities are obtained only after securing the corresponding volume regime solutions.  This was the case for the fluid regime solution, where the strong inequalities of Eq. (\ref{fluidregime}) were found to hold for the series solution of Eq. (\ref{asol0}), which was realized via the method of Eq. (\ref{perturb1}).  

In this regime, the terms involving the inverse of the higher-dimensional volume element dominate over the density and pressure components for a negative higher-
dimensional EoS parameter, $v<0$.
Here, two distinct branches of solutions are found that give rise to 3D and higher-dimensional scale factors that undergo either decelerated or accelerated expansion or contraction.  
For the aforementioned unique case of $d=1$, the 3D scale factor is found to be initially of constant finite size, however, a vanishingly small higher-dimensional volume element and density remain possible as the higher-dimensional scale factor can evolve from zero initial size, hence offering an early-time big bang scenario.

It is noted that the strong inequalities of Eq. (\ref{vacuumregime}) are in reference to the magnitude of the terms only.  Whereas in the previous section it was found that $\gamma_1>0$, in this section we find that the integration constant $\gamma_0$, with units of inverse seconds, can be of either sign for $d>1$.  We find that the allowed EoS parameter space is restricted in the volume regime, with the allowed regions determined by the signs of $\gamma_0$ and the 3D Hubble parameter.

In this regime, when the aforementioned dual strong inequalities are valid, the higher-dimensional scale factor of Eq. (\ref{bfuna}) takes on the approximate form
\beq
b(t)\sim\frac{1}{a^n(t)}\left[\int a(t)^{(dn-3)}dt\right]^{1/d}.
\eeq
The effective 4D FRW field equations can again be solved perturbatively, this time by writing the 3D scale factor as a series solution of the form
\beq\label{perturb2}
a(t)=a_0(t)+\left(\frac{\gamma_1}{\gamma_0}\right)^{(1+v)}\hspace{-.3cm}a_1(t)+...
\eeq
where, as was the case in the previous section, $a_0(t)$ is the zeroth-order solution, $a_1(t)$ is the first-order correction term, etc.  Notice that when Eq. (\ref{perturb2}) is inserted into Eqs. (\ref{rho0})-(\ref{pd0}), the density and pressure terms on the left-hand side do not yield a contribution to the zeroth-order solution as these terms' lowest-order contribution is to the first-order correction term.  Additionally, it is noted that the $\gamma_1$ terms in Eqs. (\ref{volume}) and (\ref{rhofuna}) are of order $(\gamma_1/\gamma_0)$ and will not contribute to the first-order solution as these 
are of higher-order than the $(\gamma_1/\gamma_0)^{(1+v)}$ terms, so long as $v<0$. This contrasts with the perturbative treatment of the previous section where the $\gamma_0$ terms of Eqs. (\ref{volume}) and (\ref{rhofuna}) contributed to the first-order solution as they are of order $(\gamma_0/\gamma_1)$ and therefore of the same order as that of the expansion parameter.
Following this method, we obtain the series solutions for the 3D scale factor, which take the general form
\bea\label{asolvac}
a(t)&=&a_0(c+kt)^{1/(3-d\alpha_\pm)}\nonumber\\
&\cdot&\left[1+\left(\frac{\gamma_1}{\gamma_0}\right)^{(1+v)}\kappa_\pm(c+kt)^{-\beta_\pm\delta_\pm}+...\right],
\eea
to first-order in the expansion parameter, where $a_0$, $c$, and $k$ are constants of integration.
The coefficient of the first-order correction term takes the form
\beq
\kappa_\pm=-\frac{2(1+\alpha)}{3(1+\tilde{w})}\frac{\beta_\pm^v(1-\delta_\pm)}{\delta_\pm^2(1-\beta_\pm \delta_\pm)} \frac{\tilde{a}_0^{3(1+\tilde{w})}}{a_0^{3(w-v)}k^{(1-v)}},\label{kappa-}
\eeq
where we defined the parameter combinations
\bea
\beta_\pm&\equiv&\left(\frac{n-\alpha_\pm}{3/d-\alpha_\pm}\right)\label{Y}\\
\delta_\pm&\equiv&\left(\frac{\alpha_\pm-3/(d-1)}{n-3/(d-1)}\right)(1-v).\label{Z}
\eea
Notice that in arriving at Eqs. (\ref{asolvac}) and (\ref{kappa-}) we used Eq. (\ref{a0rho0}) to exchange $\rho_0$ in favor of the constant $\tilde{a}_0$, which sets the scale for the 3D scale factor in the fluid regime (see Eq. (\ref{asol0})).  Also notice that the coefficient for the 3D scale factor, $a_0$, of Eq. (\ref{asolvac}) is left undetermined by the series solution method of this section but must be held positive for a positive 3D scale factor.

The series solutions of Eq. (\ref{asolvac}) equate to two branches of solutions in the volume regime, where the power $1/(3-d\alpha_\pm)$ takes on a negative or positive value for the top or bottom solution, respectively, where two distinct cases arise for each branch of solution.  The first case is defined by $k>0$, where the integration constant $c$ is arbitrary and only constrained by $c\geq 0$ to ensure a real 3D scale factor for all $t\geq 0$ and can therefore be set equal to zero.  
For this case, the zeroth-order solutions of Eq. (\ref{asolvac}) equate to the $D$-dimensional decelerating vacuum solutions of Chodos and Detweiler.\cite{Chodos} 
The `$\alpha_+$ solution', which corresponds to the solution of the top sign, yields decelerated contraction for $d>1$ whereas the `$\alpha_-$ solution', which corresponds to the bottom-sign solution, yields decelerated expansion for $d\geq 1$ for the 3D scale factor.

Notice that another distinct class of solutions exist if the integration constant $k<0$ and $c>0$ for a real 3D scale factor for all time. For this case, one can set $c=1$ without loss of generality. It is noted that this zeroth-order $\alpha_+$ solution equates to the $D$-dimensional accelerating vacuum solution initially realized by Levin, where the negative integration constant $k$ is
\beq\label{timeconstant}
k= -\frac{1}{t_{\scriptsize{\mbox{vol},f}}}=(3-d\alpha_\pm)H_{\scriptsize{0\pm}}(0)<0, 
\eeq
where $H_{\scriptsize{0\pm}}(0)$ is the corresponding zeroth-order Hubble parameter evaluated at $t=0$.\cite{Levin} Considering only the zeroth-order vacuum contribution, the time-contant $t_{\scriptsize{\mbox{vol},f}}$ marks the temporal end of the validity of this inflationary solution for negative $k$.

Notice that both the decelerating and accelerating classes of solutions yield either expansion or contraction for the 3D scale factor and one must chose the expanding solution as the physically relevant one. Although the $\alpha_+$ solution yields an epoch of accelerated expansion for $k<0$, the zeroth-order vacuum solution suffers from a graceful exit problem as it becomes singular at a finite time.\cite{Levin}  This problem appears nonexistent with the series solution of this section, so long as the strong inequalities of Eq. (\ref{vacuumregime}) render this regime no longer valid near $t=t_{\scriptsize{\mbox{vol},f}}$. 

Lastly, notice that for $d=1$, the zeroth-order $\alpha_+$ solution of Eq. (\ref{asolvac}) becomes \textit{constant} as $\alpha_+$ diverges for an integration constant $k$ of either sign.  To understand the dynamics of the 3D scale factor for this branch of solution for $d=1$, one must examine the first-order correction term.  This uniquely special case is closely examined later in this section and the next.

\begin{figure}[t!]
\begin{center}
	\subfigure[$\;$Plot of the power of the lowest-order $\alpha_+$ solutions versus $d$.]{\label{fig:poweralpha+}\includegraphics[width=10cm]{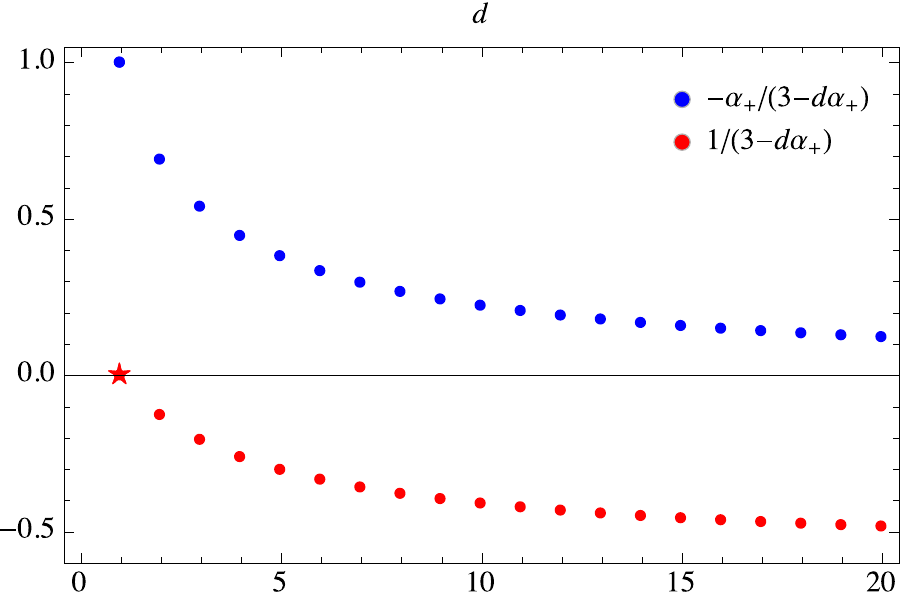}}\\
	\subfigure[$\;$Plot of the power of the lowest-order $\alpha_-$ solutions versus $d$.]{\label{fig:poweralpha-}\includegraphics[width=10cm]{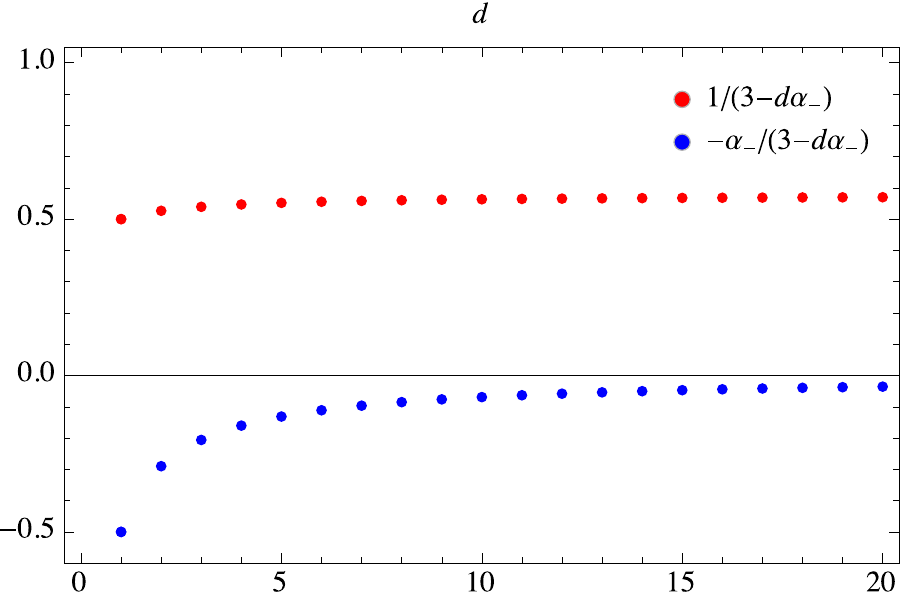}}
\end{center}
\caption{Plots of the powers $1/(3-d\alpha_\pm)$ and $-\alpha_\pm/(3-d\alpha_\pm)$, which correspond to the powers of the lowest-order solutions for $a(t)$ and $b(t)$ in the volume regime, respectively, versus $d$. Notice that the $\alpha_+$ solution exhibits decelerated contraction for $k>0$ and accelerated expansion for $k<0$ for the 3D scale factor for $d>1$, whereas the $\alpha_-$ solution exhibits decelerated expansion for $k>0$ and accelerated contraction for $k<0$ for the 3D scale factor for $d\geq 1$.  Also notice that for the unique case of $d=1$, the zeroth-order $\alpha_+$ solution for the 3D scale factor is \textit{constant} as $\alpha_+$ diverges and the corresponding power is zero. The time evolution of the 3D scale factor for this unique case is thus determined by the first-order correction term and is found to exhibit accelerated expansion for any 3D EoS parameter, $w$.  Interestingly, this special case of accelerated expansion for the 3D scale factor for $d=1$ is a member of the class of solutions that exhibit decelerated contraction for $d>1$, defined by $k>0$.}  
\end{figure}

Now, inserting Eq. (\ref{asolvac}) into Eq. (\ref{bfuna}), the higher-
dimensional scale factor takes the form
\bea\label{bsolvac}
&&b(t)=b_{\scriptsize{0}}(c+kt)^{-\alpha_\pm/(3-d\alpha_\pm)}\nonumber\\
&\cdot&\left[1-\left(\frac{\gamma_1}{\gamma_0}\right)^{(1+v)}\frac{(3-dn\delta_\pm)}{d(1-\delta_\pm)}\kappa_{\pm}(c+kt)^{-\beta_\pm\delta_\pm}+...\right],\;\;
\eea
where $\kappa_\pm$ and $\delta_\pm$ were defined in Eqs. (\ref{kappa-}) and (\ref{Z}).  Notice that the coefficient of the higher-dimensional scale factor in the volume regime is determined by the expression
\beq
b_0=\left(\frac{\gamma_0}{\beta_\pm ka_0^3}\right)^{1/d}.\label{1stob-}
\eeq
The series solutions of Eq. (\ref{bsolvac}) also equate to two branches of solutions, with two distinct cases arising for each branch.  These solutions yield either decelerated or accelerated expansion or contraction for the higher-dimensional scale factor in the volume regime.
Figures \ref{fig:poweralpha+} and \ref{fig:poweralpha-} show plots of the powers of the lowest-order solutions of $a(t)$ and $b(t)$, namely $1/(3-d\alpha_\pm)$ and $-\alpha_\pm/(3-d\alpha_\pm)$, versus the integer number of higher dimensions, $d$.  Notice that for the unique case of $d=1$, the power of the lowest-order $\alpha_+$ solution is zero, which corresponds to a constant initial scale factor.

The series solutions for the 3D and higher-dimensional scale factors, given by Eqs. (\ref{asolvac}) and (\ref{bsolvac}), respectively, generalize the $D$-dimensional decelerating vacuum solutions of Chodos and Detweiler and accelerating vacuum solutions of Levin, where the time-dependence of the zeroth-order solutions are independent of the EoS parameters $w,v$ and merely functions of the number of spacetime dimensions.\cite{Chodos,Levin}  In both of these papers, the vacuum field equations were solved without the exact relation between the 3D and higher-dimensional scale factors, hence, $b_0$ was simply an arbitrary integration constant.  In this manuscript, Eq. (\ref{bfuna}) yields a coupling between $a_0$ and $b_0$ as witnessed via Eq. (\ref{1stob-}).  By demanding that the 3D and higher-dimensional scale factors be positive and real for all values of $d$, we find two distinct possibilities for both positive and negative $\gamma_0$. 
In the next subsection, we find the relevant constraints and the allowed regions of the EoS parameter space.

Similar to that of the fluid regime, the coefficient for the higher-dimensional scale factor, given by Eq. (\ref{1stob-}), is determined collectively by the coefficient of the 3D scale factor and the integration constant $\gamma_0$.  Notice that a vanishingly small value of $\gamma_0$ can scale away the significance of Eq. (\ref{bsolvac}) in the line element, possibly offering an alternative explanation to compactification for the absence of detecting the hypothesized extra dimensions.

\subsection{Positive and real 3D and higher-dimensional scale factors}
The requirement that the 3D and higher-dimensional scale factors remain positive and real for all $d$ is equivalent to the constraints $\gamma_0>0, n>\alpha_+$ or $\gamma_0<0,n<\alpha_-$ for the case of an expanding 3D scale factor in $d>1$, where we used Eqs. (\ref{Y}), (\ref{timeconstant}), and (\ref{1stob-}) and demanded that the energy density be positive in arriving at these results.  Using  Eq. (\ref{n}), we find two possible cases of EoS parameter inequalities for both branches of expanding 3D solutions, which are given by
\bea
\gamma_0>0\;&&\mbox{and}\;\;v<1-\frac{(d-1)}{d}\left[\frac{\alpha_+-3/(d-1)}{\alpha_+-2/d}\right](1-w)\nonumber\\
&&\mbox{and}\;\;v>1-\frac{(d-1)}{d}(1-w)\;\;\;\;\;\;\;\mbox{or}\label{bposalphapm1}\\
\gamma_0<0\;&&\mbox{and}\;\;v>1-\frac{(d-1)}{d}\left[\frac{\alpha_--3/(d-1)}{\alpha_--2/d}\right](1-w)\nonumber\\
&&\mbox{and}\;\;v<1-\frac{(d-1)}{d}(1-w).\label{bposalphapm2}
\eea
\begin{figure}[h!]
\begin{center}
	\subfigure[$\;$EoS parameter space plot for the $\alpha_+$ solution with $\gamma_0>0$ and $d>1$. The green (red) region corresponds to the EoS parameters that yield \textit{accelerated expansion} (\textit{decelerated contraction}) for the 3D scale factor. ]{\label{fig:bposalpha-allA}\includegraphics[width=8cm]{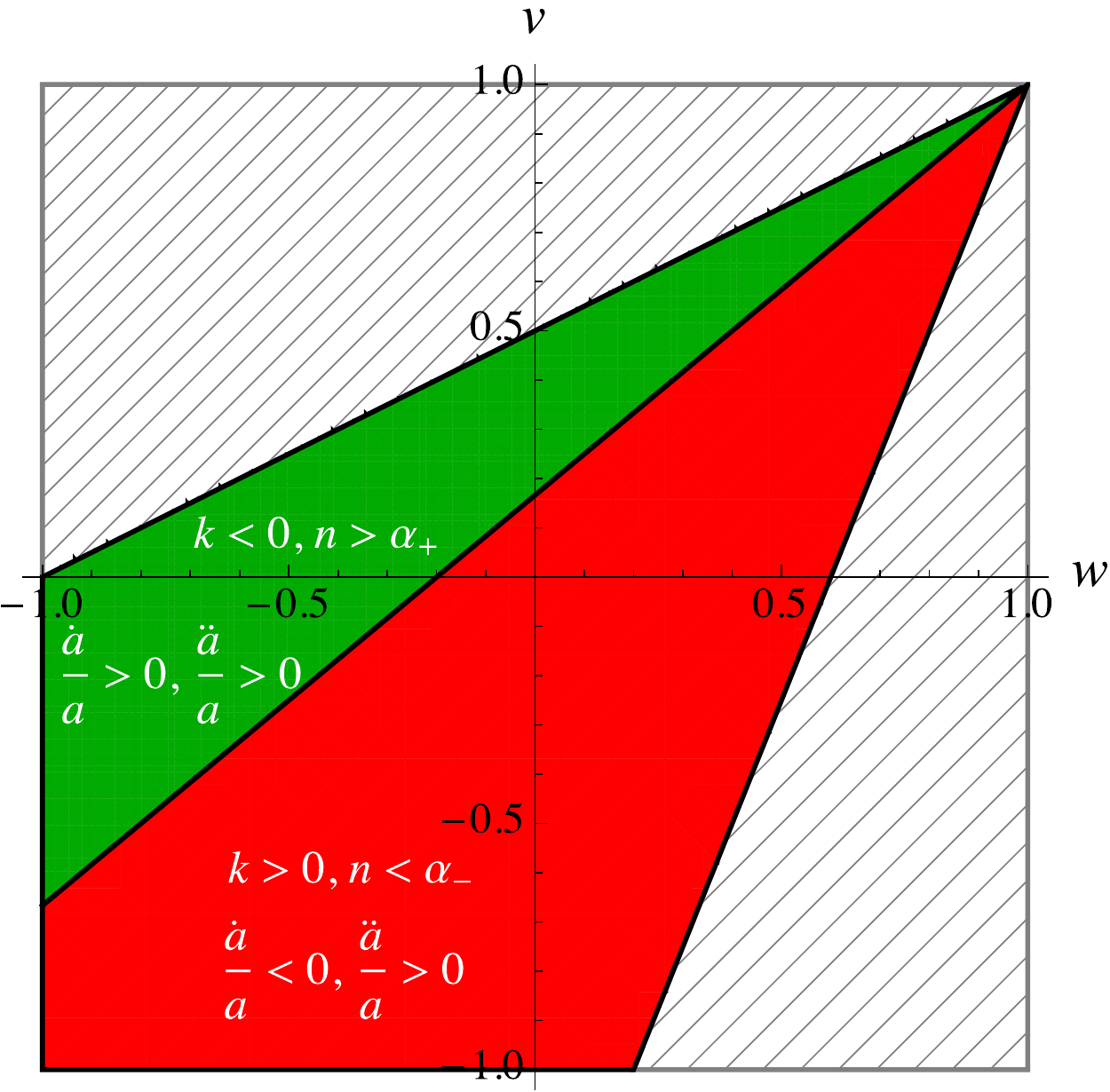}}
	\subfigure[$\;$EoS parameter space plot for the $\alpha_-$ solution with $\gamma_0<0$ and $d>1$.  The green (red) region corresponds to the EoS parameters that yield \textit{decelerated expansion} (\textit{accelerated contraction}) for the 3D scale factor.]{\label{fig:bposalpha+allA}
\includegraphics[width=8cm]{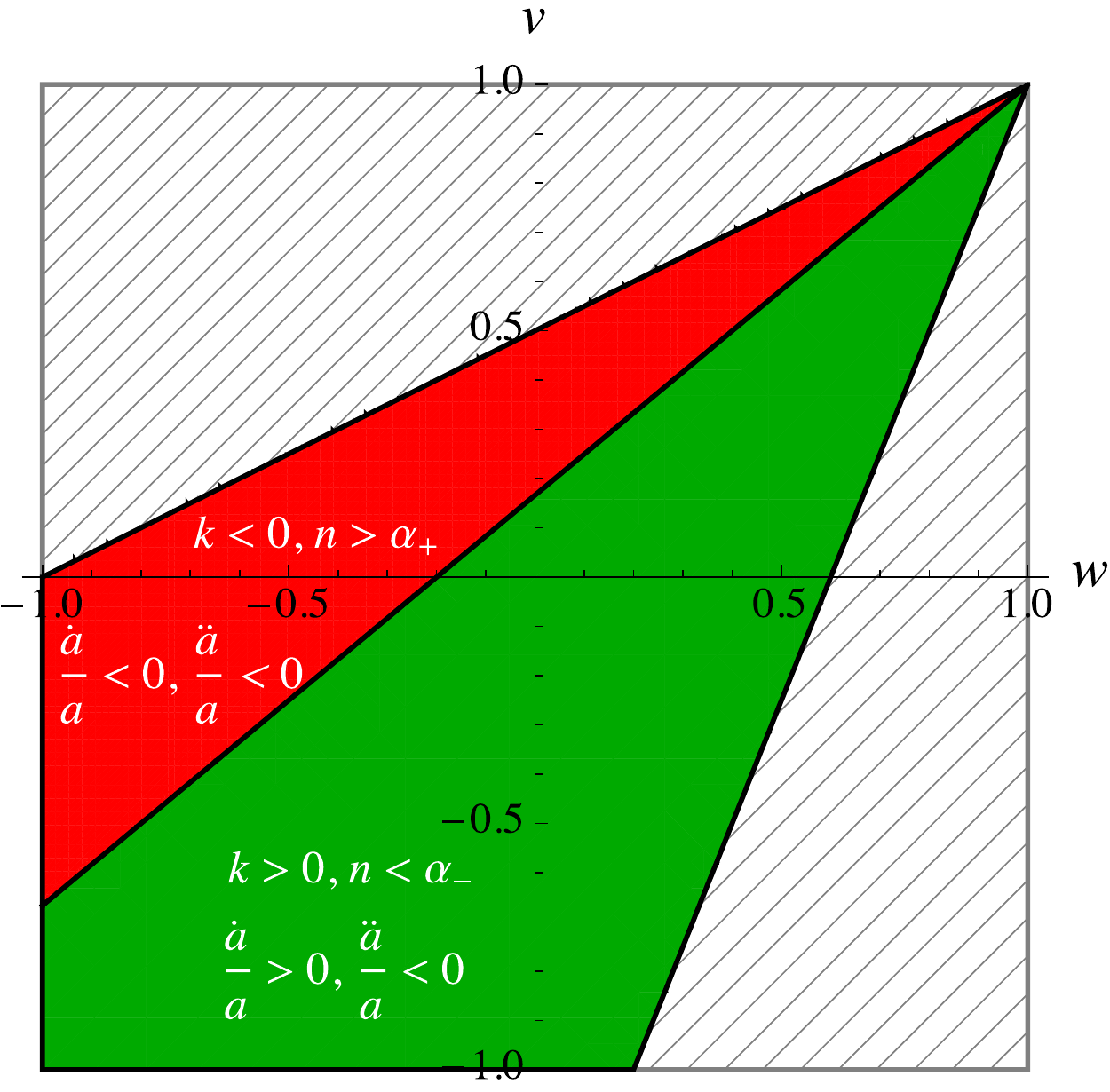}}\\
\end{center}
\caption{EoS parameter space plots of $v$ versus $w$ for the volume regime solutions.  The green and red wedge-like regions correspond to the allowed EoS parameter values that yield positive or negative values of $k$, for a given sign of $\gamma_0$, where the green (red) regions correspond to a positive (negative) 3D Hubble parameter for $d>1$.  
The hashed regions indicate the values of EoS parameters that yield a negative energy density and are not considered in this manuscript. Both of the above plots were generated for $d=6$, however, the same basic features persist for all $d>1$.  
For the $\alpha_+$ solution in $d=1$, the entire EoS parameter space with positive energy density is allowed when $\gamma_0>0$, there is no allowed region for negative $\gamma_0$.
}
\end{figure}
Figures \ref{fig:bposalpha-allA} and \ref{fig:bposalpha+allA} show plots of the allowed EoS parameter space subjected to the two cases of triple inequalities of Eqs. (\ref{bposalphapm1}) and (\ref{bposalphapm2}) for both branches of expanding 3D solutions, with these allowed regions shaded green.  In contrast, the regions indicating contracting 3D solutions are shaded red. These regions have also been constrained by the additional requirement of a positive higher-dimensional energy density.  

Notice that although the series solution method of this section generates $k$ and $a_0$ as arbitrary constants of integration, the general treatment of the effective 4D FRW field equations, which is presented in Appendix \ref{GT}, yields a prediction for their values in terms of constants and parameters (see Eqs. (\ref{k}) and (\ref{vacuumcoef})).  There we find that the $\alpha_+$ ($\alpha_-$) branch of solutions must have $\gamma_0>0$ ($\gamma_0<0$) for a real 3D scale factor.  We also find that the regions of the EoS parameter space that are shaded green yield a 3D scale factor that undergoes \textit{accelerated expansion} with $k<0$ for the $\alpha_+$ solution and \textit{decelerated expansion} with $k>0$ for the $\alpha_-$ solution.   



As previously mentioned, the $\alpha_+$ solution for the 3D scale factor for $d=1$ is constant to lowest order, hence, one must consider the first-order correction term to understand its time evolution.  
In the next subsection we show that this special case of solution is a member of the class of solutions that exhibits decelerated contraction for $d>1$, defined by $k>0$, yet gives rise to accelerated expansion for $d=1$.  For this unique case, the entire EoS parameter space with positive energy density is allowed for $\gamma_0>0$.

\subsection{The uniquely special case of the $\alpha_+$ solution in 5D}
In this subsection, we wish to closely examine the $\alpha_+$ solution in the volume regime for the special case of $d=1$.  Although this solution is found via the method outlined in Eq. (\ref{perturb2}) and is accurately described by Eq. (\ref{asolvac}) when $k>0$ for $d=1$, its regime of validity is \textit{not} dictated by Eq. (\ref{vacuumregime}) but rather when the strong inequalities
\bea\label{vacuumregime5D}
\frac{\dot{a}^2}{a^2}&\ll&\rho\sim \frac{\gamma_0}{x}\frac{\dot{a}}{a}\;\;\mbox{as}\;\;\frac{\dot{a}}{a}\ll\frac{\gamma_0}{x}\;\;\mbox{and}\nonumber\\
&&\gamma_1\ll\gamma_0\int a(t)^{(n-3)}dt
\eea
are satisfied.
In this regime, the terms in the effective 4D FRW field equations involving the inverse of the higher-dimensional volume element are found to be of the same order as the density and pressure components, but only for the $\alpha_+$ solution when $d=1$.  Here, terms involving the square of the Hubble parameter are found to be much smaller than those of the density and pressure at lowest order in this regime.  
It is interesting to note that inflation in 4D scalar field inflationary theory arises in the `slow-roll regime' when the time evolution of the scalar field is sufficiently gradual, namely when the $\dot{\phi}^2$ term is small when compared with the potential energy.\cite{GR}  In what follows, we find an early-time inflationary epoch in a regime when the Hubble parameter is small when compared with the inverse of the higher-dimensional volume element, without the need of a scalar field.

Notice that by setting $d=1$, the zeroth-order solution for the 3D scale factor, given by Eq. (\ref{asolvac}), becomes \textit{constant} as $\alpha_+$ diverges for an arbitrary integration constant $k$.  Hence, on first glance it appears that $k$ can take on either sign.  However, notice that by setting $d=1$, the zeroth-order solution for the higher-dimensional scale factor, given in Eq. (\ref{bsolvac}), becomes linear in time and exhibits either coasting expansion or contraction, depending on the sign of $k$.  To gain further insight into the dynamics of the higher-dimensional scale factor for $d=1$, one needs to look no further than the first $D$-dimensional FRW field equation, given by Eq. (\ref{density}).  In this volume regime where the strong inequalities of Eq. (\ref{vacuumregime5D}) hold, notice that this first field equation demands that the 3D and higher-dimensional Hubble parameters share the \textit{same} sign, if one demands a positive higher-dimensional density.  In this section we show that the 3D scale factor expands and so must the higher-dimensional scale factor.  Hence, we find that the $\alpha_+$ solution for $d=1$ has the additional constraint that $k>0$.  This solution is exclusively a member of the class of solutions that exhibit decelerated contraction for $d>1$, which will soon be shown to exhibit accelerated expansion for any 3D EoS parameter, $w$.

Setting $d=1$ and choosing the top sign solution, Eqs. (\ref{asolvac}) and (\ref{bsolvac}) take on the slightly simplified form
\bea\label{asolvac5D}
a(t)&=&a_0\left[1+\left(\frac{\gamma_1}{\gamma_0}\right)^{(1+v)}\hspace{-.3cm}\kappa_{+} t^{(1-v)}+...\right]\\
b(t)&=&b_0t\left[1-\left(\frac{\gamma_1}{\gamma_0}\right)^{(1+v)}\frac{3+n(1-v)}{(2-v)}\kappa_{+}t^{(1-v)}+...\right],\nonumber
\eea
where the coefficients of the first-order correction term and the higher-dimensional scale factor, originally presented in Eqs. (\ref{kappa-}) and (\ref{1stob-}), now take the form
\bea
\kappa_{+}&=&\left[\frac{2}{3(1+\tilde{w})}\right]^2\frac{\eta_1}{3(1-v)}\frac{\tilde{a}_0^{3(1+\tilde{w})}}{a_0^{3(w-v)}}\label{kappa+5D}\\
b_0&=&\frac{\gamma_0}{a_0^3}\label{b05D},
\eea
for $d=1$ where we used Eq. (\ref{parametersw}).  Notice that in arriving at Eqs. (\ref{asolvac5D}), (\ref{kappa+5D}), and (\ref{b05D}), we set $c=0$ and absorbed $k$ into our definition of $\kappa_+$ and $b_0$.  Also notice that for this $\alpha_+$ solution in $d=1$, the higher-dimensional scale factor is positive and real for $\gamma_0>0$, as can easily be seen by Eq. (\ref{b05D}).

This $\alpha_+$ solution for $d=1$ in the volume regime is special for two reasons.  As discussed earlier in this section, this solution for the 3D scale factor is \textit{constant} to lowest order, where $a_0$ corresponds to its size at $t=0$.  This approximate solution was first found in an earlier work of one the authors where the exact solution for the 3D scale factor in the 5D treatment, which was described via a product of a power of the 3D scale factor and a hypergeometric function, was expanded around a branch point.\cite{Middleton}  There it was found that the higher-dimensional EoS parameter was constrained by $v\leq 0$ when one demands that the 3D scale factor remain both positive and real.  In this work, this approximate solution in the volume regime was found through a series solution, which is valid when $v<0$.

Using the perturbative solution of Eq. (\ref{asolvac5D}), the Hubble parameter and acceleration for the 3D scale factor can easily be calculated and take the form
\bea
\frac{\dot{a}}{a}&=&\frac{\rho_0a_0^{3(v-w)}}{3\gamma_0^{(1+v)}}t^{-v}\label{vachub}\\
\frac{\ddot{a}}{a}&=&-v\cdot\frac{\rho_0a_0^{3(v-w)}}{3\gamma_0^{(1+v)}}t^{-(1+v)}\label{vacacc}
\eea
to lowest order, where we used Eq. (\ref{a0rho0}) in simplifying the expressions. As seen in Eq. (\ref{b05D}), $\gamma_0>0$ for the entire EoS parameter space for this $\alpha_+$ solution when $d=1$.  As the constants $\rho_0$ and $a_0$ are both held positive, and this approximate solution is only valid for $v<0$, Eq. (\ref{vacacc}) shows that the 3D scale factor exhibits accelerated expansion in this regime for any negative value of the higher-dimensional EoS parameter, $v$, regardless of the value of the 3D EoS parameter, $w$.  Notice that the 3D Hubble parameter increases with time, hence a positive time rate of change, whenever $v<0$ and drives the earliest expansion of the 3D scale factor from an initial constant value.  This phenomenon of accelerated expansion in the volume regime for the case of $k>0$ is unique to the 5D case as all other approximate $\alpha_+$ solutions exhibit decelerated contraction via the lowest order term (see Fig. \ref{fig:poweralpha+}).  Interestingly, this solution evades the question of naturalness, unlike the case of solutions that yield accelerated expansion for $k<0$, as there is no need to choose the expanding solution for the 3D space and the contracting solution for the higher-dimensional manifold.  For this case, both the 3D and higher-dimensional scale factors expand for $v<1$. 

In the next section, we explore \textit{when} the fluid and volume regime solutions are valid.  As each regime is characterized by a set of strong inequalities, and the time evolution of the 3D scale factor is known perturbatively, we obtain time scales that define each regime of validity.  We then show that the approximate solution for the 3D scale factor in the fluid regime can describe the early universe, and remains valid indefinitely, shortly after the even earlier volume regime solution comes into and fades out of existence.  The approximate $\alpha_+$ solution for the 3D scale factor in the volume regime for $d=1$ describes a very early-time epoch of accelerated expansion and potentially offers an alternative to scalar-field inflationary theory.

\section{A possible 5D alternative to scalar-field inflationary theory}\label{inflation}

The series solutions presented in Eqs. (\ref{asol0}) and (\ref{asolvac5D}) are valid only when the respective set of strong inequalities of Eqs. (\ref{fluidregime}) and (\ref{vacuumregime5D}) are satisfied.\footnote{In this section we limit our analysis to the special case of $d=1$, as we're interested in the regime of validity for the $\alpha_+$ solution in 5D.}  
Inserting these perturbative solutions into the corresponding strong inequalities,  we obtain time-dependent expressions of the form
\bea 
&&1\gg\frac{t_{\scriptsize{\mbox{vol},i}}}{t}\label{vacstrong1}\\
&&1\gg\left(\frac{t_{\scriptsize{\mbox{vol},i}}}{t}\right)^{(1+v)}\left(\frac{\tilde{a}_0t^{2/3(1+\tilde{w})}}{a_0}\right)^{3(1+\tilde{w})}\label{vacstrong2}\\
&&1\ll\left(\frac{t_{\scriptsize{\mbox{vol},i}}}{t}\right)\left(\frac{\tilde{a}_0t^{2/3(1+\tilde{w})}}{a_0}\right)^{(3-n)},\label{fluidstrong}
\eea
where the first two inequalities collectively define the regime of validity for the $d=1$ volume regime $\alpha_+$ solution. Notice that the set of strong inequalities of Eq. (\ref{fluidregime}) can be shown to be effectively equivalent, with the help of the fluid regime solution in Eq. (\ref{asol0}), and reduce to the lone expression given by Eq. (\ref{fluidstrong}), which solely defines the regime of validity for the fluid solution. Also notice that in arriving at Eqs. (\ref{vacstrong2}) and (\ref{fluidstrong}) we dropped overall multiplicative factors that depend on EoS parameters as here we're interested in obtaining the order of magnitude of the relevant time scales.

In arriving at the above strong inequalities, we defined the fundamental time scale
\beq\label{plancktime}
t_{\scriptsize{\mbox{vol},i}}\equiv\frac{\gamma_1}{\gamma_0}a_0^{3-n},
\eeq
which is present in each expression and proves to be a useful definition.  Notice that once the time becomes much larger than this fundamental time scale, the strong inequality of Eq. (\ref{vacstrong1}) is consequently satisfied and the approximate solution of the volume regime, given by Eq. (\ref{asolvac5D}), becomes valid and is consequently `turned on'.  Eqs. (\ref{vacstrong2}) and (\ref{fluidstrong}) feature two competing time-dependent quantities, with the second term a ratio of the fluid to volume regime solutions at lowest order.   It is interesting to note that the regimes of validity of the volume and fluid regime solutions are ultimately determined by the values of the constants $t_{\scriptsize{\mbox{vol},i}}$ and $a_0$, in addition to the EoS parameters, $w$ and $v$.  As the coefficient for the 3D scale factor, $\tilde{a}_0$, can be chosen to normalize the fluid regime solution to yield $a(t_0)=1$ today, we set 
\bea\label{m}
\tilde{a}_0&=&t_0^{-2/3(1+\tilde{w})}\nonumber\\
a_0&\equiv&10^{m}\nonumber\\
t_{\scriptsize\mbox{vol},i}&\equiv&10^\tau\mbox{s},
\eea
where $t_0\sim10^{17}\mbox{s}$ and we parameterized the constant initial 3D scale factor and the fundamental time constant.  As the time constant $t_{\scriptsize{\mbox{vol},i}}$ sets the scale for when the volume regime solution becomes valid, we will eventually set this constant equal to the Planck time, $t_{\scriptsize{\mbox{vol},i}}=t_P\sim 10^{-43}\mbox{s}$, for now we leave it as a free parameter.  Notice that at the Planck time, the 3D scale factor is predicted by flat 4D FRW cosmology to have a value $a_{\scriptsize{\mbox{4D}}}(t_P)\sim 10^{-30}$ for $w=1/3$ in a radiation-dominated epoch.  Therefore, we anticipate that $m$ should have a value near $m\sim -30$ if the constant initial scale factor $a_0$ is to align with the value predicted by standard 4D FRW cosmology for the 3D scale factor when evaluated at the Planck time. 

Upon adopting the parameterizations of Eq. (\ref{m}), Eqs. (\ref{vacstrong1})-(\ref{fluidstrong}) can be rewritten into the form
\bea
&&1\ll\left(\frac{t}{t_{\scriptsize{\mbox{vol},i}}}\right)\hspace{1.05cm}\mbox{where}\;\;t_{\scriptsize{\mbox{vol},i}}\equiv 10^\tau\mbox{s}\label{vs0}\\
&&1\gg\left(\frac{t}{t_{\scriptsize{\mbox{vol},f}}}\right)^{(1-v)}\;\;\mbox{where}\;\;t_{\scriptsize{\mbox{vol},f}}\equiv 10^q\mbox{s}\label{vs}\\
&&1\ll\left(\frac{t}{t_{\scriptsize{\mbox{flu},i}}}\right)^{-(1+\alpha)}\;\mbox{where}\;\;t_{\scriptsize{\mbox{flu},i}}\equiv 10^p\mbox{s},\label{fs}
\eea
where the powers of the aforementioned time scales are determined by the algebraic expressions
\bea
q&=&\frac{2}{(1-v)}\left[\frac{3}{2}(1+\tilde{w})-\frac{(\tau-17)}{m}\right]m+\tau\label{q}\\
p&=&\frac{\alpha}{(1+\alpha)}\left[\frac{3}{2}(1+\tilde{w})-\frac{(\tau-17)}{m}\right]m+\tau\label{p}\;,
\eea
with $\tau$ a free parameter.   Again, once the time becomes much larger than the fundamental time constant defined in Eq. (\ref{plancktime}), Eq. (\ref{vs0}) becomes satisfied and the volume regime is turned on.  Notice that the right-hand side of Eq. (\ref{vs}) also increases with time as $v<1$.  Hence, the $\alpha_+$ volume regime solution is `turned off' once the strong inequality of Eq. (\ref{vs}) is no longer satisfied.  As the $\alpha_+$ solution of the volume regime exhibits accelerated expansion for \textit{any} value of the 3D EoS parameter, $w$, as long as the higher-dimensional EoS parameter is negative, and is found to turn on and off as dictated by Eqs. (\ref{vs0}) and (\ref{vs}), this solution potentially describes an early-time inflationary epoch for the 3D scale factor when radiation is the dominant energy density component. 
Additionally, we find that the right-hand side of Eq. (\ref{fs}) increases with time as $\alpha<-1$ for $d=1$ for the EoS parameters in the domain $-1\leq w,v<1$ and a positive energy density, where $\alpha$ was defined in Eq. (\ref{parametersw}).  Hence, the series solution of the fluid regime, given by Eq. (\ref{asol0}), is only `turned on' once the strong inequality of Eq. (\ref{fs}) becomes satisfied.  As the fluid regime is defined by only one unique strong inequality, the fluid regime solution is never turned off once turned on and offers a relatively late-time behavior for the 3D scale factor.

It is interesting to note that the special case $\tau=q=p$ occurs if 
\beq\label{mequal}
\frac{3}{2}(1+\tilde{w})=\frac{(\tau-17)}{m},
\eeq 
which is equivalent to the more revealing equality $a_0=\tilde{a}_0t_{\scriptsize\mbox{vol},i}^{2/3(1+\tilde{w})}$.  Hence, if Eq. (\ref{mequal}) is satisfied there is no regime of validity for the 5D volume solution as the volume regime turns off when turning on as the relevant time scales marking these phenomena are made equivalent, which is easily witnessed by examination of Eqs. (\ref{vs0}) and (\ref{vs}) with $q=\tau$.  Additionally, the time scale indicating the beginning of the fluid regime equals that of the volume regime as $p=\tau$.  For this case, the scale factor evolving as the fluid regime solution at $t_{\scriptsize\mbox{vol},i}$ must equate to the value of the constant initial scale factor of the volume regime, if Eq. (\ref{mequal}) is to be satisfied.  Equation (\ref{mequal}) equates to a temporal boundary condition between solutions when the time interval for the volume regime is collapsed zero.

\subsection{Lower bound on the constant initial 3D scale factor}

Here we are interested in the case where the fluid regime begins after the volume regime ends with the volume regime ending only after initially beginning, which is equivalent to the set of inequalities $p>q>\tau$.  
These time ordering constraints are collectively met when the power of the initial constant scale factor of the volume regime is larger than a threshold value, given by
\beq\label{mth}
m>m_{\scriptsize\mbox{th}}\equiv\frac{2(\tau-17)}{3(1+\tilde{w})},
\eeq
which is equivalent to the inequality $a_0>a_{\scriptsize\mbox{0,th}}\equiv\tilde{a}_0t_{\scriptsize\mbox{vol},i}^{2/3(1+\tilde{w})}$.  
Namely, the value of the constant initial scale factor of the volume regime, $a_0$, must be larger than what the scale factor would be at $t_{\scriptsize\mbox{vol},i}$ if it were evolving as a fluid regime solution, if time ordering is obeyed.

\begin{figure}[!t]
\begin{center}
\includegraphics[width=15cm]{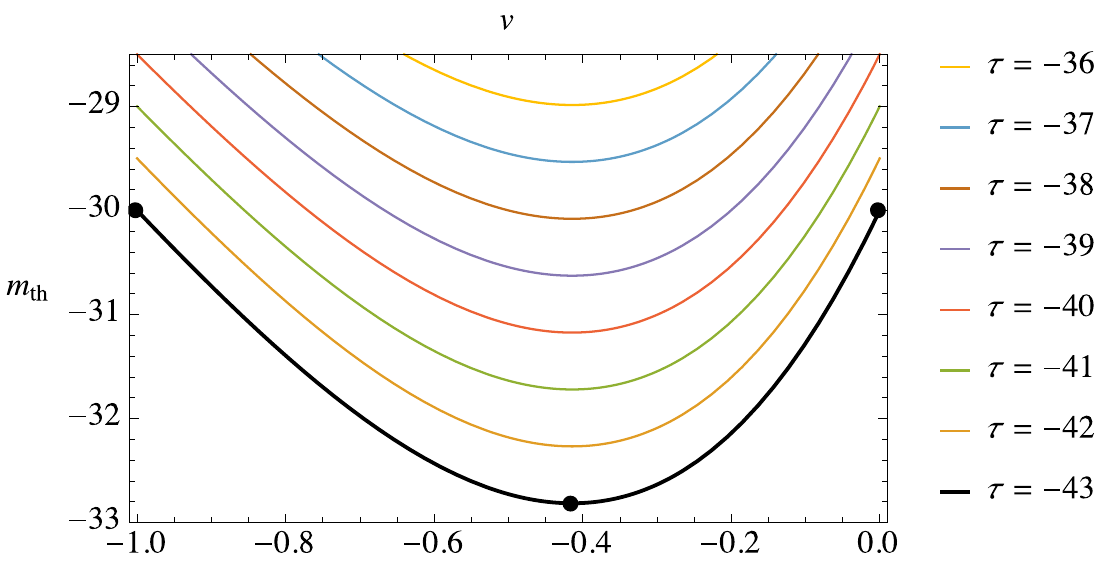}
\caption{Plot of the threshold power $m_{\scriptsize\mbox{th}}$ versus $v$ for $w=1/3$, which corresponds to a radiation-dominated universe for a traceless 4D stress energy tensor.  This threshold power defines a lower bound on the constant initial scale factor, $a_0=10^m$, for the volume regime solution as $m>m_{\scriptsize\mbox{th}}$ if time ordering is to be obeyed. 
Notice that if the fundamental time constant is set equal to the Planck time, which is equivalent to setting $\tau=-43$, then $m_{\scriptsize{\mbox{th}}}=-30$ at both $v=0$ and $v=-1$.  This equates to the lower bound on the constant initial scale factor matching the value predicted by standard 4D FRW cosmology at the Planck time for a radiation-dominated universe. For $-1<v<0$, the threshold power is smaller than this value with an absolute minimum of $m_{\scriptsize\mbox{th,min}}\simeq -32.8$ at $v_{\scriptsize\mbox{min}}\simeq -0.41$, nearly three orders of magnitude smaller than the value predicted by 4D FRW cosmology.}
\label{fig:mth}
\end{center}
\end{figure}

Figure \ref{fig:mth} shows a plot of the threshold power, $m_{\scriptsize\mbox{th}}$, defined in Eq. (\ref{mth}), versus the higher-dimensional EoS parameter, $v$, for several values of $\tau$, where we set $w=1/3$ for a radiation-dominated universe with a traceless 4D stress-energy tensor. Notice that the shape of the $m_{\scriptsize\mbox{th}}$ curve is independent of the chosen value of $\tau$.  Also notice that this threshold power matches the value predicted by standard 4D FRW cosmology when evaluated at the beginning of the volume regime, at both $v=0$ and $v=-1$ for a radiation-dominated epoch described by $w=1/3$.  This is due to the fact that 
\bea
\tilde{w}|_{v=0}&=&1/3\;\;\;\mbox{for all}\;w\\
\tilde{w}|_{v=-1}&=&w,
\eea
hence, $a_{\scriptsize\mbox{0,th}}=\tilde{a}_0t_{\scriptsize\mbox{vol},i}^{2/3(1+w)}$ at $v=0$ and $v=-1$. Notice that if the higher-dimensional EoS parameter takes on values in the range $-1<v<0$, the threshold power is smaller than that predicted by 4D FRW cosmology when evaluated at the beginning of the volume regime.  The minimum of the threshold power, $m_{\scriptsize\mbox{th,min}}$, can be found by setting the $v$-derivative of $m_{\scriptsize\mbox{th}}$ equal to zero, which yields 
\bea
m_{\scriptsize\mbox{th,min}}&=&\frac{(\tau-17)}{(3w-2)+2\sqrt{3(1-w)}}\;\;\mbox{at}\\
v_{\scriptsize\mbox{min}}&=&1-\sqrt{3(1-w)}\;\;\;\mbox{for}\;-1\leq v\leq 0.\nonumber
\eea
For a radiation-dominated universe, we find $m_{\scriptsize\mbox{th,min}}=(\tau-17)(1+2\sqrt{2})/7$ at $v_{\scriptsize\mbox{min}}=1-\sqrt{2}\simeq -0.41$ for all $\tau$. It is noted that for the remainder of this section we limit our discussion to $w=1/3$, describing a radiation-dominated universe with a traceless 4D stress-energy tensor.

\subsection{Lower and upper bounds on the end of the volume regime and the beginning of the fluid regime}

\begin{figure}[t!]
\begin{center}
\subfigure[$\;$
Notice that for the $m=-30$ curve, $q$ has a maximum value of $q_{\scriptsize{\mbox{max}}}=-35.5$ at $v_{\scriptsize{\mbox{max}}}=-1/3$.  This places an upper bound on the end of the volume regime, where the corresponding solution exhibits accelerated expansion.]{\label{fig:powersq}\includegraphics[width=15cm]{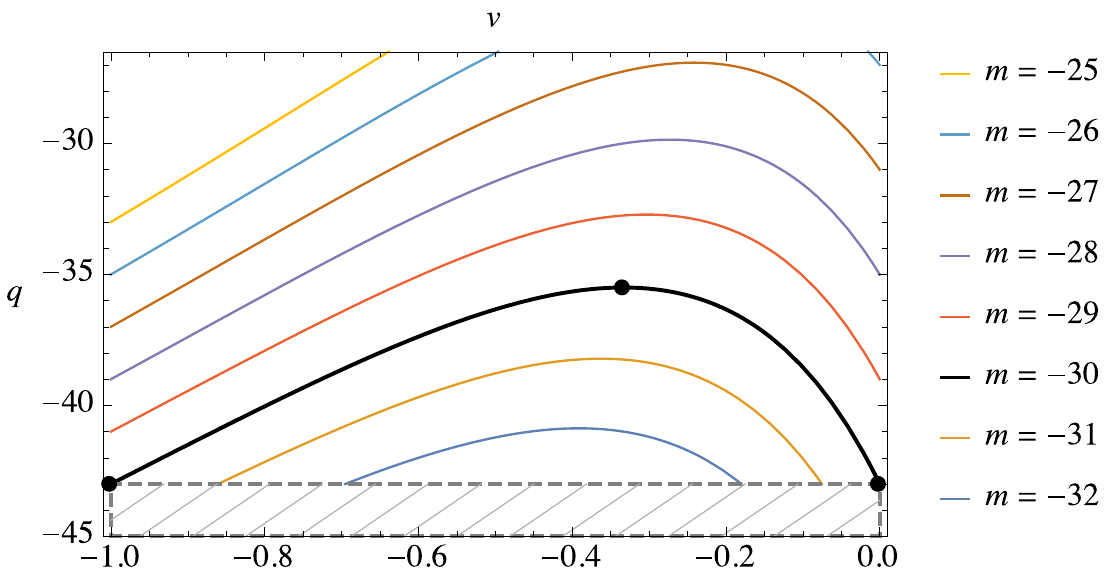}}\\
\subfigure[$\;$
Notice that for the $m=-30$ curve, $p$ has a maximum value of $p_{\scriptsize{\mbox{max}}}=-13$ at $v_{\scriptsize{\mbox{max}}}=-1$, where the slope is zero.  This places an upper bound on the beginning of the fluid regime, where the corresponding solution closely mimics that of 4D FRW cosmology.]{\label{fig:powersp}\includegraphics[width=15cm]{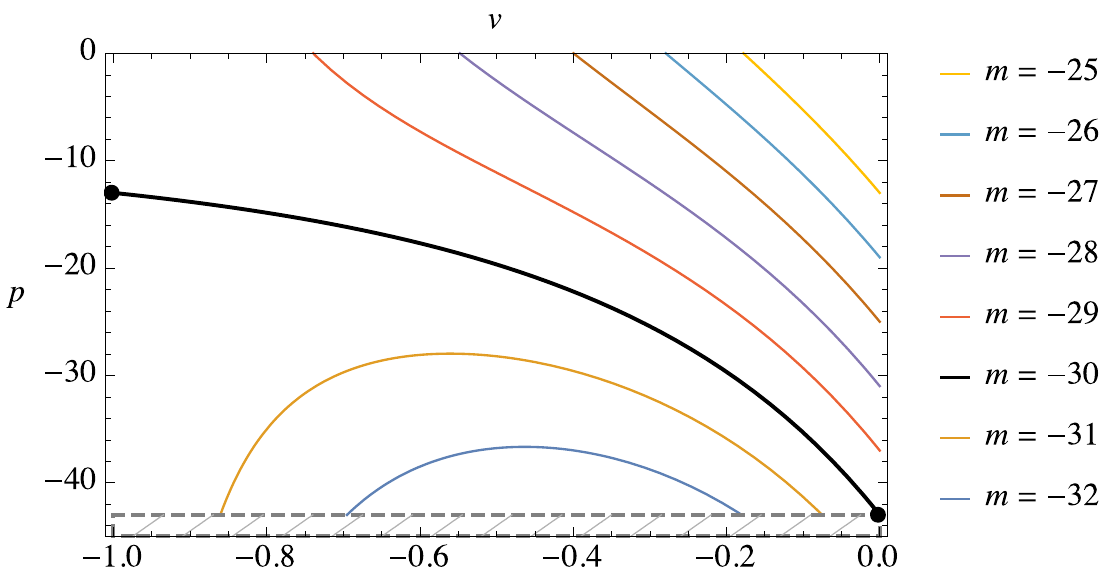}}
\end{center}
\caption{Plots of the powers $q$ vs $v$ and $p$ vs $v$ for $w=1/3$, which corresponds to a radiation-dominated universe for a traceless 4D stress energy tensor.  Here we set $\tau=-43$, which sets the time constant that determines when the volume regime solution becomes valid, $t_{\scriptsize{\mbox{vol},i}}$, equal to the Planck time.  These powers define the time scales $t_{\scriptsize{\mbox{vol},f}}\equiv 10^q\mbox{s}$ and $t_{\scriptsize{\mbox{flu},i}}\equiv 10^p\mbox{s}$ characterizing when the $\alpha_+$ volume regime solution `turns off' and when the fluid regime solution `turns on', respectively. 
Notice that for a traceless 5D stress energy tensor, equating to the EoS parameter relation $v=1-3w$, similar ranges for $q$ and $p$ are found, however, the shape of the $p$ vs $v$ curve is markedly different.}
\end{figure}

Although the power of the constant initial scale factor of the volume regime is only bounded from below as $m>m_{\scriptsize\mbox{th}}$, the power of the time scale describing the end of the volume regime, $q$, is bounded from \textit{both} below and above for a given $m$ value. Figure \ref{fig:powersq} shows a plot of this power, as defined in Eqs. (\ref{vs}) and (\ref{q}), as a function of the higher-dimensional EoS parameter, $v$, for several values of $m$, where we set 
$\tau=-43$ aligning the fundamental time constant with the Planck time.  
The hashed region equates to $q<\tau$, which indicates the region where time ordering is not obeyed and is therefore not considered in this manuscript.  Notice that the lower bound on the power of the time scale describing the end of the volume regime, $q_{\scriptsize\mbox{min}}$, has two distinct possibilities depending upon the chosen value of $m$ for a given $\tau$, which are characterized by 
\bea\label{qmin}
&&q_{\scriptsize\mbox{min}}=\tau\;\;\mbox{for}\;\;\frac{1}{7}(1+2\sqrt{2})(\tau-17)<m\leq\frac{1}{2}(\tau-17)\;\;\mbox{or}\nonumber\\
&&q_{\scriptsize\mbox{min}}=q\rvert_{v=-1}=2m+17\;\;\;\;\mbox{for}\;\;\;\;m\geq\frac{1}{2}(\tau-17).
\eea 
Notice that for the unique case of $m=(\tau-17)/2$, the lower bound occurs at both $v=-1$ and $v=0$. Additionally, $q_{\scriptsize\mbox{min}}$ for $m=(\tau-17)/2$ satisfies both expressions in Eq. ({\ref{qmin}) as $m=(\tau-17)/2$ marks the boundary between these two distinct cases of $q_{\scriptsize{\mbox{min}}}$.  

The upper bound on the power of the time scale describing the end of the volume regime, $q_{\scriptsize\mbox{max}}$, can be found by setting the $v$-derivative of $q$ equal to zero, which yields
\bea\label{qmax}
q_{\scriptsize\mbox{max}}&=&-\frac{1}{4m}\left[(\tau-17)^2+2(\tau-17)m-7m^2\right]+\tau\;\;\mbox{at}\nonumber\\
v_{\scriptsize\mbox{max}}&=&\frac{(\tau-17)-3m}{(\tau-17)+m}\;\;\mbox{for}\;-1\leq v\leq 0.
\eea
Hence, for a chosen value of $m$ for a given $\tau$, the time scale describing the end of the volume regime is constrained with its power residing within the range $q_{\scriptsize\mbox{min}}\leq q\leq q_{\scriptsize\mbox{max}}$, where $q_{\scriptsize\mbox{min}}$ and $q_{\scriptsize\mbox{max}}$ are determined by Eqs. (\ref{qmin}) and (\ref{qmax}), respectively.

The power of the time scale describing the beginning of the fluid regime, $p$, is bounded from below and \textit{in some cases} from above, depending on the chosen value of $m$ for a given $\tau$.  Figure \ref{fig:powersp} shows a plot of this power, as defined in Eqs. (\ref{fs}) and (\ref{p}), as a function of the higher-dimensional EoS parameter, $v$, for several values of $m$ where we again set $\tau=-43$.  The hashed region equates to $p<\tau$, which is not considered here as time ordering is not obeyed in this region.  The lower bound on the power of the time scale describing the beginning of the fluid regime, $p_{\scriptsize\mbox{min}}$, also has two distinct possibilities depending upon the chosen value of $m$ for a given $\tau$, which are characterized by
\bea
&&p_{\scriptsize\mbox{min}}=\tau\;\mbox{for}\;\frac{1}{7}(1+2\sqrt{2})(\tau-17)<m\leq\frac{1}{2}(\tau-17)\;\mbox{or}\nonumber\\
&&p_{\scriptsize\mbox{min}}=p\rvert_{v=0}=3(2m+17)-2\tau\;\mbox{for}\;m\geq\frac{1}{2}(\tau-17).
\nonumber\\\label{pmin2}
\eea 
Notice that for the unique case of $m=(\tau-17)/2$, $p_{\scriptsize\mbox{min}}$ satisfies both expressions in Eqs. (\ref{pmin2}) as $m=(\tau-17)/2$ marks the boundary between these two distinct cases.

The upper bound on the power of the time scale describing the beginning of the fluid regime, $p_{\scriptsize\mbox{max}}$ also has two distinct possibilities depending upon the chosen value of $m$ for a given $\tau$, where only one of these possibilities is finite.  The finite upper bound is found by setting the $v$-derivative of $p$ equal to zero, which yields 
\bea\label{pmax}
p_{\scriptsize\mbox{max}}=-\left(\frac{3-v_{\scriptsize\mbox{max}}}{1-v_{\scriptsize\mbox{max}}}\right)^2\left(\frac{v_{\scriptsize\mbox{max}}^2-2v_{\scriptsize\mbox{max}}-1}{v_{\scriptsize\mbox{max}}^2-6v_{\scriptsize\mbox{max}}+1}\right)m+\tau&&\nonumber\\
\mbox{for}\;\;\frac{1}{7}(1+2\sqrt{2})(\tau-17)<m\leq\frac{1}{2}(\tau-17)\;\;\mbox{at}&&\nonumber\\
\frac{(\tau-17)}{m}=\left(\frac{v_{\scriptsize\mbox{max}}^2-v_{\scriptsize\mbox{max}}+2}{1-v_{\scriptsize\mbox{max}}}\right)\;\;\;\;\;\;\;\mbox{for}\;-1\leq v\leq 0&&\nonumber\\
+\left(\frac{3-v_{\scriptsize\mbox{max}}}{1-v_{\scriptsize\mbox{max}}}\right)(1+v_{\scriptsize\mbox{max}})\left(\frac{v_{\scriptsize\mbox{max}}^2-2v_{\scriptsize\mbox{max}}-1}{v_{\scriptsize\mbox{max}}^2-6v_{\scriptsize\mbox{max}}+1}\right).\nonumber\\&&
\eea
Notice that the above set of algebraic equations are coupled, where $v_{\scriptsize\mbox{max}}$ must satisfy both expressions when yielding the local maximum of $p$.  Interestingly, the largest \textit{finite} $p_{\scriptsize\mbox{max}}$ occurs at $v=-1$ for $m=(\tau-17)/2$ with $p_{\scriptsize\mbox{max}}$ diverging to either positive or negative infinity at $v=-1$ for $m>(\tau-17)/2$ or $m<(\tau-17)/2$, respectively.

Hence, for a chosen value of $m$ for a given $\tau$, the time scale describing the beginning of the fluid regime is constrained with its power residing within the range $p_{\scriptsize\mbox{min}}\leq p\leq p_{\scriptsize\mbox{max}}$ for $(\tau-17)(1+2\sqrt{2})/7<m\leq(\tau-17)/2$, where $p_{\scriptsize\mbox{min}}$ is determined by the first expression of Eq. (\ref{pmin2}) and $p_{\scriptsize\mbox{max}}$ is determined by Eq. (\ref{pmax}).  For the distinct case of $m>(\tau-17)/2$, the time scale describing the beginning of the fluid regime is \textit{only} constrained from below with its power $p\geq p_{\scriptsize\mbox{min}}$, where $p_{\scriptsize\mbox{min}}$ is determined by the second expression of Eq. (\ref{pmin2}).

It is interesting to note that for $\tau=-43$ and $m=-30$, which correspond to the constant initial scale factor, $a_0$, aligning with the value of the 3D scale factor predicted by standard 4D FRW cosmology at the Planck time, the time scales determining the end of the volume regime and the beginning of the fluid regime take on maximum values given by
\bea
t_{\scriptsize{\mbox{vol},f,\mbox{max}}}\sim 10^{-35.5}\mbox{s}\nonumber\\
t_{\scriptsize{\mbox{flu},i,\mbox{max}}}\sim 10^{-13}\mbox{s},
\eea
at $v_{\scriptsize{\mbox{max}}}=-1/3$ and $v_{\scriptsize{\mbox{max}}}=-1$, respectively.
These time scales, curiously, match remarkably well with the times predicted for the unification of the electromagnetic and weak force, or electroweak (ew) force, and the unification of the strong and electroweak force in the Grand Unified Theory (GUT) given by $t_{\scriptsize{\mbox{ew}}}\sim 10^{-12}\mbox{s}$ and $t_{\scriptsize{\mbox{GUT}}}\sim 10^{-36}\mbox{s}$, respectively. \cite{ryden} 

\subsection{Lower and upper bounds on $\gamma_1/\gamma_0$}

Finally, we discuss the behavior of the higher-dimensional scale factor, $b(t)$, in 5D in the volume and fluid regimes.  Notice that as the $\alpha_+$ solution for the 3D scale factor in the volume regime undergoes accelerated expansion for $-1\leq v<0$ and $w=1/3$, which corresponds to a radiation-dominated epoch in the very early universe with a traceless 4D stress energy tensor, the higher-dimensional scale factor expands as $b(t)\sim t$ to lowest order.  After the volume regime ends and the fluid regime begins, notice that $n>0$ for $w=1/3$ and $-1\leq v<0$ (see Eq.(\ref{n}) or Fig.\ref{fig:all}), which corresponds to a scenario where the higher-dimensional scale factor undergoes dynamical compactification (see Eq. (\ref{moham})).  Later, during a pressureless matter-dominated epoch of the fluid regime, $w=0$, the higher-dimensional scale factor undergoes either dynamical compactification or decelerated expansion, depending on 
whether $-1\leq v<-1/2$ or $v>-1/2$.  In the late universe, when dark energy is the dominant fluid component and the 3D scale factor undergoes accelerated expansion with $-1\leq\tilde{w}<-1/3$, the higher-dimensional scale factor once again experiences decelerated expansion as $n<0$.  It is currently unclear whether the dynamical compactification of the higher-dimensional scale factor during the early fluid regime is enough to dominate over the expansion it experiences during the early volume and late fluid regimes, when the 3D scale factor exhibits accelerated expansion, to yield an extra dimension that has contracted to a small enough size that the universe becomes effectively four dimensional.

Vanishingly small values of $\gamma_1$ and $\gamma_0$ can scale away the overall significance of the higher-dimensional scale factor in the line element, given by Eq. (\ref{metric}), offering an alternative explanation for the extra dimension's absence from observation, regardless of whether the higher-dimensional scale factor undergoes dynamical compactification or expansion.  Notice that the integration constants $\gamma_1$ and $\gamma_0$ act as overall scales for the higher-dimensional scale factor, $b(t)$, in a given regime (see Eqs. (\ref{tildeb0}) and (\ref{b05D})), and seem to be left undetermined by the theory.  However, the ratio of the integration constants $\gamma_1/\gamma_0$ is constrained and is bounded from below and above for a chosen value of $m$ for a given $\tau$.  Equation (\ref{plancktime}) yields an expression for this ratio.  Setting $\tau=-43$ and $m=-30$, which corresponds to the constant initial scale factor, $a_0$, aligning with the value of the 3D scale factor predicted by standard 4D FRW cosmology at the Planck time, the ratio of these integration constants is constrained to reside within the range
\beq\label{gams}
10^{17}\mbox{s}\leq \frac{\gamma_1}{\gamma_0}\leq 10^{47}\mbox{s}\;\;\mbox{with}\;\;\frac{\gamma_1}{\gamma_0}\sim 10^{32}\mbox{s}\;\;\mbox{at}\;\;q_{\scriptsize{\mbox{max}}},
\eeq
where, again, $\gamma_1$ is a unitless numerical scale whereas $\gamma_0$ has units of inverse seconds.  Again, although this ratio is constrained via Eq. (\ref{gams}), each integration constant is left undetermined by the theory and can seemingly be chosen to take on vanishingly small values, scaling away the significance of the higher-dimensional scale factor and offering an alternative explanation for its absence from observation.

\section{Conclusion} \label{conclusion}

Here we examined the time evolution of the $D=d+4$ dimensional Einstein field equations subjected to a flat Robertson-Walker metric, where the scale factors for the 3D and higher-dimensional spaces were allowed to evolve at different rates in the general case.  We chose the higher-dimensional stress energy tensor to be that of a perfect fluid, where the 3D and higher-dimensional pressures were allowed to differ from one another.  Following in a manner analogous to that of standard 4D FRW cosmology, we adopted two equations of state linearly relating the 3D and higher-dimensional pressures to the density.  We arrived at an exact expression for the higher-dimensional scale factor, written solely in terms of a function of the 3D scale factor.  This expression allowed us to decouple the higher-dimensional field equations and arrive at a set of effective 4D FRW field equations, written exclusively in terms of the 3D scale factor. The higher-dimensional scale factor is determined by two competing terms, each characterized by an associated integration constant, whose dominance partially defines either regime of validity for the two series solutions.  

We then examined the effective 4D FRW field equations in the general case and arrived at the exact solution, which equates to the third integral of these non-linear differential equations. This resultant expression relates a function of the 3D scale factor, namely a product of an integral function of the 3D scale factor and a hypergeometric function, to the time.  This expression cannot, in general, be inverted to yield an analytical expression for the 3D scale factor as a function of time.  This exact expression can, however, be expanded near the three singularities of the hypergeometric function, 
which ultimately yields three series solutions in two unique regimes.  Equivalently, one can solve the effective 4D FRW field equations perturbatively in each of these two distinct volume and fluid regimes. Each regime is characterized by a set of strong inequalities that ultimately define the time interval when the corresponding series solution is valid.  We withheld the exact treatment until Appendix \ref{GT} and presented the perturbative treatment within the body of this paper as the latter proves to yield a simpler path to 
our main results.  

The fluid regime solution was found to be valid when a set of strong inequalities are obeyed; these are fundamentally characterized by when the Hubble parameter is much larger than the inverse of the higher-dimensional volume element.  This regime emerges when the higher-dimensional scale factor is dominated by the first of its two terms and is consequently proportional to a power of the 3D scale factor.  This, interestingly, equates to a generalized treatment of dynamical compactification first studied by Mohammedi.\cite{Mohammedi}  There, $n$ was an arbitrary power held constant to ensure contraction of the higher-dimensional scale factor, whereas here dynamical compactification occurs naturally for a limited range of 3D and higher-dimensional EoS parameters, $w$ and $v$.  The solution for the 3D scale factor in this fluid regime was found to have precisely the same functional form as that of standard 4D FRW cosmology to lowest order, with the EoS parameter, $w$, replaced with the effective EoS parameter, $\tilde{w}$.  It was found that this solution offers a late-time epoch of accelerated expansion, but only for a limited range of the 3D and higher-dimensional EoS parameters, $w$ and $v$.  Interestingly, neither of the regions of the EoS parameter space plot that yield dynamical compactification overlap with the region that yields accelerated expansion for the 3D scale factor.  Thus, one concludes that if the 3D scale factor is undergoing accelerated expansion in the fluid regime, then the higher-dimensional scale factor cannot be simultaneously undergoing dynamical compactification.

The volume regime solutions were found in the regime where the terms involving the inverse of the higher-dimensional volume element dominate over the density and pressure components for $d>1$.  
This regime emerges when the higher-dimensional scale factor is dominated by the second of its two terms and is proportional to an integral function of the 3D scale factor. 
The 3D scale factor in this volume regime is described by series solutions where the lowest order contribution equates to a generalized treatment of the $D$-dimensional accelerating vacuum solutions of Levin and decelerating vacuum solutions of Chodos and Detweiler.\cite{Levin,Chodos}  These series solutions equate to two distinct branches of solutions, where the 3D and higher-dimensional scale factors can exhibit decelerated or accelerated expansion or contraction.  We showed that either case of each branch of solutions arises for a limited range of 3D and higher-dimensional EoS parameters, $w$ and $v$.  For the uniquely special case of $d=1$, the 3D scale factor takes the form of a constant to lowest order for one branch of solution; the time evolution of the 3D scale factor is hence determined by the first-order correction term at lowest order.  This solution was found to obey a different set of strong inequalities, which equate to a time when the Hubble parameter was much smaller than the inverse of the higher-dimensional volume element.  This solution was found to give rise to an early-time accelerated expansion for \textit{any} value of the 3D EoS parameter, $w$, including during a radiation-dominated epoch, so long as the higher-dimensional EoS parameter is negative, $v<0$. Hence, this model offers a possible alternative to scalar-field inflationary theory.

Lastly, we examined the strong inequalities that define the regimes of validity for the series solutions of the volume and fluid regimes for $d=1$, which corresponds to a 5D spacetime.  
The volume regime is ultimately determined by two unique strong inequalities, where the corresponding solution was shown to turn on and then off in the very early universe, hence possibly offering a natural exit from an inflationary epoch.  This very early-time solution is proceeded by the relatively late-time fluid regime solution, whose regime of validity is ultimately determined by only one unique strong inequality.  This fluid regime solution was shown to turn on after the volume regime solution turns off and consequently remains on indefinitely, yielding a late-time solution that closely mimics that predicted by 4D FRW cosmology.  We showed that this time ordering is obeyed so long as
the constant initial 3D scale factor of the volume regime is larger than a predicted threshold value.  This threshold value matches the value of the 3D scale factor of the fluid regime, which matches that predicted by standard 4D FRW cosmology at $v=0$ and $v=-1$, when evaluated at the start of the volume regime.
Further, we found that the time scales marking the end of the volume regime and the beginning of the fluid regime are constrained to reside within a predicted range, with both time scales bounded from below and above.  By aligning the fundamental time constant, which marks the beginning of the volume regime, with the Planck time and the corresponding constant initial 3D scale factor with the value predicted by standard 4D FRW cosmology at this time, we showed that the time scales determining the end of the volume regime and the beginning of fluid regime take on maximum values of $t_{\scriptsize{\mbox{vol},f,\mbox{max}}}\sim 10^{-35.5}\mbox{s}$ and $t_{\scriptsize{\mbox{flu},i,\mbox{max}}}\sim 10^{-13}\mbox{s}$.  It was noted that these time scales match remarkably well with the times predicted for the unification of the strong and electroweak force and the unification of the electromagnetic and weak force, respectively, when the universe was a small fraction of a second old. 

As the solutions presented in Sections \ref{fluid} and \ref{vacuum1} describing the aforementioned phenomena are of a perturbative nature, it is left to show that the exact solution of the effective 4D FRW field equations, presented in Appendix \ref{GT} via Eq. (\ref{solution}), can in fact generate the number of $e$-folds necessary to solve the horizon and flatness problem, if the model presented in this manuscript is to truly offer a higher-dimensional alternative to scalar-field inflation theory.  This examination surely can be done through numerical methods.  Additional work is also required to see if this model is capable of generating the irregularities necessary to lead to the formation of structure.

Additionally, this model offers a possible alternative explanation for the lack of observation of the higher dimensions from that offered by dynamical compactification.  The overall size of the higher-dimensional scale factor is dictated by, in part, the constants $\gamma_0$ and $\gamma_1$ for the volume and fluid regimes, respectfully.  Although the ratio of these constants is constrained, their individual numeric values appear to be left undetermined by the theory.  As previously suggested, vanishingly small values for $\gamma_0$ and $\gamma_1$ can scale away the significance of the higher-dimensional scale factor from the line element.

\appendix
\section{General treatment of the effective 4D FRW field equations}\label{GT}

Finally, we present an exact treatment of the 4D effective FRW field equations for a single fluid component.  Remarkably, these equations can be integrated thrice to yield an exact solution involving the 3D scale factor.  This resulting solution relates a product of a power of an integral of the 3D scale factor and a hypergeometric function to the time, and cannot, in general, be inverted to obtain $a=a(t)$.  Rather, this exact expression can be expanded in different regimes to yield approximate solutions for the 3D scale factor, which equate to the series solutions found in the fluid and volume regimes presented earlier in this manuscript.

Using Eq. (\ref{effective}) to eliminate $\rho$ and $\tilde{p}$ from Eqs. (\ref{rho0}) and (\ref{p0}), factoring and performing some algebra, we obtain an expression dictating the behavior of the 3D scale factor of the form
\bea\label{eom}
&&\frac{d}{dt}\left(\ln\left(\frac{\dot{a}}{a}+\frac{2}{3(1+\tilde{w})}\frac{(1-\beta_\pm)}{\alpha\beta_\pm}\frac{\gamma_0}{x}\right)\right)+\frac{3}{2}(1+\tilde{w})\frac{\dot{a}}{a}\nonumber\\
&&+\left(v-\frac{(1+\alpha)}{\alpha}\frac{(1-\beta_\pm)}{\beta_\pm}\right)\frac{\gamma_0}{x}=0\;\;\mbox{for}\;\;\tilde{w}\neq -1,
\eea
where $\alpha$ and $\beta_\pm$ were previously defined in Eqs. (\ref{parametersw}) and (\ref{Y}) and we used the parameter relationship of Eq. (\ref{parametersw}).  
Eq. (\ref{eom}) can be integrated with the help of the relation
\beq\label{gammax}
\frac{\gamma_0}{x}=\frac{d}{dt}\left(\ln\left(xa^{(dn-3)}\right)\right),
\eeq
where we used Eq. (\ref{volume}) in generating this expression.  Substituting Eq. (\ref{gammax}) into Eq. (\ref{eom}), integrating and then rearranging, we arrive at the first integral of the field equations, which takes the form
\bea\label{firstintegral}
c_0&=&\left(xa^{(dn-3)}\right)^{[v-(2+\alpha)(1-\beta_\pm)/\alpha\beta_\pm]}\nonumber\\
&\cdot&\frac{d}{dt}\left(a^{3(1+\tilde{w})/2}\left(xa^{(dn-3)}\right)^{(1-\beta_\pm)/\alpha\beta_\pm}\right),
\eea
where $c_0\geq 0$, but is otherwise an arbitrary constant of integration. Notice that the value of this constant can be chosen to normalize the 3D scale factor at the present time.  

Motivated by the suggestive form of Eq. (\ref{firstintegral}), we define the integral function
\beq\label{f}
f\equiv xa^{(dn-3)}=\gamma_1+\gamma_0\int a^{(dn-3)} dt,
\eeq
where we used Eq. (\ref{volume}) in arriving at the second equality.  As is obvious from Eq. (\ref{f}), the defined function is determined by two constants and an integral of a power of the 3D scale factor.  This effectively equates the 3D scale factor to a power of the time derivative of the defined function, which can easily be seen by differentiating Eq. (\ref{f}).  The adoption of Eq. (\ref{f}) consequently allows for the integration of Eq. (\ref{firstintegral}), which will become apparent shortly.

Now, using Eq. (\ref{f}) to eliminate the 3D scale factor from Eq. (\ref{firstintegral}) in favor of our defined function and then subsequently integrating, we arrive at a second integral of the field equations of the form
\beq\label{integral}
\left(\dot{f}f^{(1-\beta_\pm)/\beta_\pm}\right)^{(1+\alpha)/\alpha}\hspace{-.2cm}=-c_0(1+\alpha)\frac{\gamma_0^{1/\alpha}}{\delta_\pm}\left(f^{-\delta_\pm}-c_1\right),
\eeq
where the parameter $\delta_\pm$ was previously defined in Eq. (\ref{Z}) and $c_1$ is a second integration constant.  As will be shown later in this section, the resultant exact expression can be approximated in two limiting regimes, which are found to equate to those of the fluid and volume regimes of Sec. \ref{fluid} and \ref{vacuum1}.  When these approximate solutions are then compared with the series solutions of the fluid and volume regimes, we find agreement when
\beq\label{c1gamma1}
c_1=\gamma_1^{-\delta_\pm}.
\eeq
With the benefit of hindsight, we use this identification throughout the rest of this section. 

Now, Eq. (\ref{integral}) can easily be separated and then integrated.  Doing this, we obtain the \textit{exact} expression, which relates a product of a power of an integral of the 3D scale factor and a hypergeometric function to the time, which takes the form
\bea\label{solution}
f^{1/\beta_\pm}&\cdot&\;_2F_1\left(\frac{\alpha}{(1+\alpha)},-\frac{1}{\beta_\pm \delta_\pm} ;1-\frac{1}{\beta_\pm \delta_\pm};\frac{\gamma_1^{\delta_\pm}}{f^{\delta_\pm}}\right)\nonumber\\
&=&(c+kt),
\eea
where $c$ is a third integration constant and we defined the parameter
\beq\label{k}
k\equiv \frac{\gamma_0}{\beta_\pm}\left(c_0\gamma_1^{-\delta_\pm}\frac{(1+\alpha)}{\gamma_0\delta_\pm}\right)^{\alpha/(1+\alpha)}.
\eeq
Notice that Eq. (\ref{solution}) cannot, in general, be inverted to yield the 3D scale factor as an analytical function of time.  To gain further insight into the behavior of the 3D scale factor, whose time evolution is described by the exact expression of Eq. (\ref{solution}), one must employ either numerical techniques or analytical approximations.  Also notice that ${}_2F_1(\alpha,\beta;\gamma;z)$ is a hypergeometric function that can be defined in terms of a hypergeometric series of the form
\beq\label{series}
_2F_1(a,b;c; z)=1+\frac{ab}{c}\;z+\frac{a(a+1)b(b+1)}{2c(c+1)}\;z^2+...
\eeq
In general, the hypergeometric series has singularities at $z=0,1,\;\mbox{and}\;\infty$, with a branch point at $z=1$ (see \cite{table,table2}).  In the following subsections, we wish to explore the approximate behavior of our exact expression near these singularities.  Using hypergeometric transformations and the hypergeometric series given by Eq. (\ref{series}), we expand Eq. (\ref{solution}) near each singularity and obtain perturbative solutions for the integral function $f$.  Taking a time derivative consequently yields the approximate behavior for the 3D scale factor near the $z=0,1,\;\mbox{and}\;\infty$ singularities.  These approximate solutions equate to the series solutions of the fluid and volume regimes, which were previously found in Sections \ref{fluid} and \ref{vacuum1}.

\subsection{$f^{\delta_\pm}\gg \gamma_1^{\delta_\pm}$}\label{fgg}

Notice that when $(\gamma_1/f)^{\delta_\pm}$ is small one can write the hypergeometric function of Eq. (\ref{solution}) as a hypergeometric series, which is useful for obtaining the approximate behavior of Eq. (\ref{solution}) in this limiting regime.  Applying the series expansion of Eq. (\ref{series}) to Eq. (\ref{solution}) and performing some algebra, we obtain a non-linear expression of the form
\beq\label{h2}
\kappa_2 h^2+h=(c+kt)^{\ell},
\eeq
where we kept terms up to order $o(h^2)$ in generating the above expression and defined the quantities
\bea\label{hell1}
\ell&\equiv&-\beta_\pm \delta_\pm\nonumber\\
h&\equiv&f^{-\delta_\pm}\nonumber\\
\kappa_2&\equiv&\frac{\alpha}{(1+\alpha)}\frac{\ell}{(1+\ell)}\gamma_1^{\delta_\pm}.
\eea
As we are interested in the time evolution of the 3D scale factor for small $h$, Eq. (\ref{h2}) can be solved perturbatively by writing $h$ as a series solution of the form
\beq
h=h_0+h_1+...\;,
\eeq
where $h_0$ is the solution to the linear approximation of Eq. (\ref{h2}), $h_1$ is the first-order correction term, etc. Using this method, we obtain an approximate solution for $h$ of the form
\beq\label{h}
h=(c+kt)^{\ell}\left[1-\kappa_2(c+kt)^{\ell}+...\right].
\eeq
Now, using the definition of $h$ given in Eq. (\ref{hell1}), we obtain the approximate solution for $f$, which can then be differentiated to arrive at a perturbative solution for the 3D scale factor.  We obtain an approximate solution of the form
\bea\label{vacuum}
a(t)&=&a_0(c+kt)^{1/(3-d\alpha_\pm)}\nonumber\\
&\cdot&\left[1-\kappa_2\frac{(1-1/\delta_\pm)}{(dn-3)}(c+kt)^{-\beta_\pm\delta_\pm}+...\right],
\eea
where we used Eq. (\ref{f}) in obtaining this result.  The coefficient for the 3D scale factor is determined by
\beq\label{vacuumcoef}
a_0\equiv\left(c_0\gamma_1^{-\delta_\pm}\frac{(1+\alpha)}{\gamma_0\delta_\pm}\right)^{2/3(1+\tilde{w})(1+\alpha)}
\eeq
where we used Eqs. (\ref{f}), (\ref{k}), and (\ref{hell1}) in arriving at these expressions.  The series solutions of Eq. (\ref{vacuum}) equate to two distinct branches of solutions in this small $(\gamma_1/f)^{\delta_\pm}$ regime.

It is interesting to note that $k$ and $a_0$ were found to be arbitrary constants of integration via the series solution method of Section \ref{vacuum1}, but are determined by constants and parameters via Eqs. (\ref{k}) and (\ref{vacuumcoef}) through the general treatment presented here.  Now, as $c_0$ and $\gamma_1$ have each been found to be strictly positive, we arrive at the additional constraint $(1+\alpha)/\gamma_0\delta_\pm>0$ when one demands a real 3D scale factor.  Employing Eqs. (\ref{alphapm}), (\ref{parametersw}), and (\ref{Z}), one can easily show that $(1+\alpha)/\delta_+>0$ and $(1+\alpha)/\delta_-<0$ for all allowed values of the EoS parameters that yield a positive energy density as $-1<\tilde{w}$ for $-1< w,v<1$.  Hence, we find two possible cases that yield a real 3D scale factor described by
\bea\label{inequal21}
&&\gamma_0>0\;\;\mbox{and}\;\;\frac{(1+\alpha)}{\delta_+}>0\;\;\mbox{or}\nonumber\\
&&\gamma_0<0\;\;\mbox{and}\;\;\frac{(1+\alpha)}{\delta_-}<0.
\eea
For the $\alpha_+$ ($\alpha_-$) solution of Eq. (\ref{vacuum}) we find that $\gamma_0>0$ ($\gamma_0<0$).  

Notice that when either of the above cases are met we obtain not only a real but also positive value for $a_0$. This contrasts with the fact that the parameter $k$ can be positive or negative, whose sign depends on the values of the EoS parameters $w,v$. Upon examination of Eq. (\ref{k}), we find that 
\bea\label{posnegk}
&&\mbox{for}\;\;\gamma_0>0\;,\;k<0\;\;\mbox{when}\;\;n>\alpha_+\;\;\mbox{or}\nonumber\\
&&\mbox{for}\;\;\gamma_0<0\;,\;k>0\;\;\mbox{when}\;\;n<\alpha_-
\eea
for an expanding 3D scale factor, where we used Eq. (\ref{Y}) and demanded a positive energy density.  Figures \ref{fig:bposalpha-allA} and \ref{fig:bposalpha+allA} show plots of the EoS parameter space for each of the cases that yield a real 3D scale factor and a parameter $k$ of either sign, dictated by Eqs. (\ref{inequal21}) and (\ref{posnegk}), respectively, when subjected to the additional requirement of a positive energy density.  The regions of the EoS parameter space that are shaded green yield a 3D scale factor that undergoes \textit{accelerated expansion} for the $\alpha_+$ solution and \textit{decelerated expansion} for the $\alpha_-$ solution.


Now inserting Eq. (\ref{vacuum}) into Eq. (\ref{bfuna}), the higher-
dimensional scale factor takes the form
\bea\label{vacuumb}
b(t)&=&b_0(c+kt)^{-\alpha_\pm/(3-d\alpha_\pm)}\nonumber\\
&\cdot&\left[1+\kappa_2\frac{(n-3/d\delta_\pm)}{(dn-3)}(c+kt)^{-\beta_\pm\delta_\pm}+...\right],\;\;
\eea
where we defined the coefficient for the higher-dimensional scale factor through the relation
\beq\label{vacuumcoefb}
b_0\equiv\left(\frac{\gamma_0}{\beta_\pm ka_0^3}\right)^{1/d}.
\eeq
Notice that these approximate solutions for the 3D and higher-dimensional scale factors, given by Eqs. (\ref{vacuum}) and (\ref{vacuumb}), match the series solutions of the volume regime, given by Eqs. (\ref{asolvac}) and (\ref{bsolvac}), respectively.  Also, the coefficient for the higher-dimensional scale factor, given by Eq. (\ref{vacuumcoefb}), precisely matches that found in the volume regime given by Eq. (\ref{1stob-}).  Notice, however, that the method employed in this section yields the additional condition for $k$ and the coefficient of the 3D scale factor, $a_0$, which are determined by Eqs. (\ref{k}) and (\ref{vacuumcoef}), respectively.

\subsection{$f^{\delta_\pm}\ll \gamma_1^{\delta_\pm}$}\label{z0b}

In this subsection, we wish to study the behavior of Eq. (\ref{solution}) when $(f/\gamma_1)^{\delta_\pm}$ is small.  This small $(f/\gamma_1)^{\delta_\pm}$ behavior can be investigated by first applying a hypergeometric transformation of the form
\bea\label{trans1}
&&_2F_1(a,b; c ;z)\nonumber\\
&&=\frac{\Gamma(c)\Gamma(b-a)}{\Gamma(b)\Gamma(c-a)}\;(-z)^{-a}\;_2F_1(a,1-c+a;1-b+a; 1/z)\nonumber\\
&&+\frac{\Gamma(c)\Gamma(a-b)}{\Gamma(a)\Gamma(c-b)}\;(-z)^{-b}\;_2F_1(b,1-c+b; 1-a+b; 1/z)\nonumber\\
\eea
to Eq. (\ref{solution}) and then expanding as a hypergeometric series for small $1/z$.  Applying Eq. (\ref{trans1}) to Eq. (\ref{solution}) and performing some algebra, we obtain the expression
\beq\label{hyper2}
f^{\delta_\pm/\ell}{}\cdot\;_2F_1\left(\frac{\alpha}{(1+\alpha)},\frac{1}{\ell};\frac{(1+\ell)}{\ell};\frac{f^{\delta_\pm}}{\gamma_1^{\delta_\pm}}\right)=(c+kt),
\eeq
where we redefined the quantities $k$ and $\ell$ of the previous subsection as
\bea\label{hell2}
k&\equiv& \frac{\gamma_0\delta_\pm}{\ell}\left(-c_0\frac{(1+\alpha)}{\gamma_0\delta_\pm}\right)^{\alpha/(1+\alpha)}\nonumber\\
\ell&\equiv&\beta_\pm \delta_\pm\left(1+\frac{\alpha\beta_\pm \delta_\pm}{(1+\alpha)}\right)^{-1}
\eea
in this subsection.
Notice that in obtaining Eq. (\ref{hyper2}), the second term of the transformation equation multiplied by $f^{1/\beta_\pm}$ reduces merely to a function of the EoS parameters and the integration constant $\gamma_1$.  This term, which is void of the 3D scale factor, can be absorbed into the integration constant $c$, effectively shifting the size of the 3D scale factor at a given $t$.

We now proceed by following a process similar to that presented in the previous subsection.  Employing the hypergeometric series given by Eq. (\ref{series}) for small $(f/\gamma_1)^{\delta_\pm}$ and again keeping terms up to and including order $o(h^2)$, we again obtain an equation of the form given by Eq. (\ref{h2}), with the parameters $h$ and $\kappa_2$ redefined as
\bea\label{hell2also}
h&\equiv&f^{\delta_\pm}\nonumber\\
\kappa_2&\equiv&\frac{\alpha}{(1+\alpha)}\frac{\ell}{(1+\ell)}\gamma_1^{-\delta_\pm},
\eea
where $k$ and $\ell$ were defined in Eq. (\ref{hell2}). Using the solution for $h$, which was presented in Eq. (\ref{h}) with the parameters specified in Eqs. (\ref{hell2}) and (\ref{hell2also}), we obtain an expression for $f$ and then differentiate.  We obtain an approximate solution for the 3D scale factor of the form
\bea\label{vacuumalso}
a(t)&=&a_0(c+kt)^{1/(3-d\alpha_\mp)}\nonumber\\
&\cdot&\left[1-\kappa_2\frac{(1+1/\delta_\pm)}{(dn-3)}(c+kt)^{-\beta_\mp\delta_\mp}+...\right],
\eea
where the coefficient for the 3D scale factor is determined by
\beq\label{vacuumcoef2}
a_0\equiv\left(-c_0\frac{(1+\alpha)}{\gamma_0\delta_\pm}\right)^{2/3(1+\tilde{w})(1+\alpha)},
\eeq
where we used Eqs. (\ref{f}) and (\ref{hell2}) in arriving at these expressions.  Notice that the series solutions of Eq. (\ref{vacuumalso}) equate to two distinct branches of solutions in this small $(f/\gamma_1)^{\delta_\pm}$ regime and are equivalent to those presented in Eq. (\ref{vacuum}).  Upon close examination of the approximate solutions of this subsection and the last, one finds that one can arrive directly at Eqs. (\ref{hell2})-(\ref{vacuumcoef2}) from Eqs. (\ref{k}), (\ref{hell1}), (\ref{vacuum}) and (\ref{vacuumcoef}) by making the parameter substitutions 
\bea\label{transdb}
&&\delta_\pm\rightarrow -\delta_\pm=\delta_\mp\nonumber\\
&&\beta_\pm\rightarrow\beta_\pm\left(1+\frac{\alpha\beta_\pm \delta_\pm}{(1+\alpha)}\right)^{-1}=\beta_\mp
\eea
and by setting $\gamma_1=1$.
Hence, the top sign solution of Eq. (\ref{vacuum}) is equivalent to the bottom sign solution of Eq. (\ref{vacuumalso}) and vice versa when one sets $\gamma_1=1$.  The approximate solutions for the 3D scale factor near the $z=0$ and $z=\infty$ singularities equate to two different branches of equivalent solutions.

\subsection{$f\sim\gamma_1$}\label{z1}
Lastly, we explore the behavior of Eq. (\ref{solution}) in the vicinity of the $f/\gamma_1\sim 1$ branch point.  In this section, we employ another hypergeometric transformation formula of the form
\bea\label{trans}
_2F_1(a,b;c ;z)=\frac{\Gamma(c)\Gamma(a+b-c)}{\Gamma(a)\Gamma(b)}\;(1-z)^{c-a-b}&&\nonumber\\
\cdot\;_2F_1(c-a, c-b ;c-a-b+1;1-z)&&\nonumber\\
+\frac{\Gamma(c)\Gamma(c-a-b)}{\Gamma(c-a)\Gamma(c-b)}\;_2F_1(a,b ;a+b-c+1;1-z).&&
\eea
By applying this transformation to the hypergeometric function in Eq. (\ref{solution}) and performing some algebra, one can show that Eq. (\ref{solution}) takes the form
\bea\label{123}
(c+kt)&=&h^{1/\ell}\left(1-h\right)^{-1/\beta_\pm\delta_\pm}\\
&\cdot&\;_2F_1\left(1\; ,1-\frac{1}{\beta_\pm \delta_\pm}\left(1+\frac{\alpha\beta_\pm \delta_\pm}{(1+\alpha)}\right);\frac{(1+\ell)}{\ell};h\right),\nonumber
\eea 
where we again redefined the parameters $k,\ell,$ and $h$ of the previous subsections as
\bea\label{hell3}
k&\equiv&\gamma_1^{-1/\beta_\mp}\left(c_0^\alpha\frac{\gamma_0\delta_\pm}{(1+\alpha)}\right)^{1/(1+\alpha)}\nonumber\\
\ell&\equiv&(1+\alpha)\nonumber\\
h&\equiv&1-\frac{\gamma_1^{\delta_\pm}}{f^{\delta_\pm}}
\eea
in this subsection, where we used Eq. (\ref{transdb}) in arriving at these expressions.
As was the case in the previous subsection, the second term of the transformation equation when multiplied by $f^{1/\beta_\pm}$ reduces to a function of the EoS parameters and the integration constant $\gamma_1$, which can be shown through use of the hypergeometric series of Eq. (\ref{series}).  This term, which is void of the 3D scale factor, can be absorbed into the integration constant $c$, which effectively shifts the size of the scale factor at a given $t$.

Now applying the series expansion of Eq. (\ref{series}) for small $h$ to the hypergeometric function in Eq. (\ref{123}) and performing some algebra, we again arrive at an expression of the form given by Eq. (\ref{h2}).  In arriving at this expression we again kept terms up to order $o(h^2)$ and redefined the parameter 
\beq\label{hell3also}
\kappa_2\equiv\frac{\ell}{(1+\ell)}\left(1+\frac{1}{\beta_\pm \delta_\pm}\right),
\eeq
where $k,\ell,$ and $h$ were previously defined in Eq. (\ref{hell3}). Using the solution for $h$, which was presented in Eq. (\ref{h}) with the parameters specified in Eqs. (\ref{hell3}) and (\ref{hell3also}), we obtain an expression for $f$ and then differentiate.  We obtain an approximate solution of the form
\bea\label{sol1}
a(t)&=&\tilde{a}_0(c+kt)^{2/3(1+\tilde{w})}\\
&\cdot&\left[1-\frac{2v/\delta_\pm}{3(1+\tilde{w})(2+\alpha)}(c+kt)^{(1+\alpha)}+...\right],\nonumber
\eea
where the coefficient for the 3D scale factor in this limiting regime is determined by
\beq\label{fluidcoef}
\tilde{a}_0\equiv\left[\gamma_1^{-(1-\beta_\mp)/\beta_\mp}\hspace{-.1cm}\left(c_0\frac{(1+\alpha)}{\gamma_0\delta_\pm}\right)^{\alpha/(1+\alpha)}\right]^{1/(dn-3)},
\eeq
Notice that in the fluid regime, the parameters $k$ and $\tilde{a}_0$, given by Eqs. (\ref{hell3}) and (\ref{fluidcoef}), are real when either of the cases of Eq. (\ref{inequal21}) are satisfied.  This result is precisely equivalent to that for real values of $k$ and $a_0$ in the volume regime.  However, in the volume regime we found that $k$ could take on positive or negative values, with the sign of $k$ determined by the inequalities of Eq. (\ref{posnegk}).  In the fluid regime, we find that $k$ is strictly positive, which can easily be witnessed through examination of Eq. (\ref{hell3}) as $\gamma_1>0$. Notice that for $k>0$, the integration constant $c$ is arbitrary, only constrained by $c\geq 0$ to ensure a real 3D scale factor for all $t\geq 0$, and can therefore be set equal to zero.

Now, inserting Eq. (\ref{sol1}) into Eq. (\ref{bfuna}), the higher-dimensional scale factor takes the form
\bea\label{bfunasol1}
b(t)&=&\tilde{b}_0(c+kt)^{-2n/3(1+\tilde{w})}\\
&\cdot&\left[1+\frac{1}{d\delta_\pm}\left[1+\frac{2v\cdot dn}{3(1+\tilde{w})(2+\alpha)}\right](c+kt)^{(1+\alpha)}+...\right],\nonumber
\eea
where the coefficient for the higher-dimensional scale factor in this limiting regime is determined by
\beq
\tilde{b}_0\equiv\frac{\gamma_1^{1/d}}{\tilde{a}_0^n}.
\eeq
Notice that in calculating Eq. (\ref{bfunasol1}), the integral term in Eq. (\ref{bfuna}) contributes to the first-order correction term and must be included.  Also notice that the approximate solutions for the 3D and higher-dimensional scale factors given by Eqs. (\ref{sol1}) and (\ref{bfunasol1}) can be compared to the series solutions obtained in the fluid regime, given by Eqs. (\ref{asol0}) and (\ref{bfunasol0}).  
One finds perfect agreement between these solutions when the integration constant $c=0$, which can be done without loss of generality as $k>0$ in the fluid regime.

\end{document}